\documentclass[]{aastex7}
\pdfoutput=1 
\usepackage{natbib}
\let\tablenum\relax
\usepackage{siunitx}
\usepackage{chemformula}
\usepackage{gensymb}
\usepackage{amssymb}
\usepackage{subcaption}
\usepackage{graphicx}
\usepackage[super]{nth}
\usepackage[version=4]{mhchem}
\usepackage{tablefootnote}

\begin{document}

\title{The Role of Tectonic Luck in Long-Term Habitability of Abiotic Earth-like Planets}

\author[0000-0002-0508-857X]{Brandon Park Coy}
\email{bpcoy@uchicago.edu}
\correspondingauthor{Brandon Park Coy}
\email{bpcoy@uchicago.edu}

\affiliation{Department of the Geophysical Sciences, University of Chicago, Chicago, Illinois, United States}

\author[0000-0002-1426-1186]{Edwin S. Kite}
\email{kite@uchicago.edu}
\affiliation{Department of the Geophysical Sciences, University of Chicago, Chicago, Illinois, United States}
\author[0000-0001-9289-4416]{R.J. Graham}
\email{arejaygraham@uchicago.edu}
\affiliation{Department of the Geophysical Sciences, University of Chicago, Chicago, Illinois, United States}

\begin{abstract}

Carbonate-silicate weathering feedback is thought to stabilize Earth's climate on geologic timescales. If climate warms, faster mineral dissolution and increased rainfall speed up weathering, increasing \ce{CO2} drawdown and opposing the initial warming. Limits to where this feedback might operate on terrestrial exoplanets with \ce{N2-O2-CO2-H2O} atmospheres 
are used to define the `habitable zone'---the range of orbits around a star where liquid water can be stable on a planet's surface. However, the impacts  on long-term habitability of randomly varying volcanic outgassing, tectonic collisions, and tectonic parameters (e.g., number of continental plates, size of plates, plate velocity) remain poorly understood. In this work, we present an idealized and broadly-applicable quasi-2D model of the long-term climate stability of abiotic Earth-twins.  The model tracks atmospheric \ce{CO2} as `disks' collide, promoting uplift and supplying new weatherable minerals through erosion.  Without resupply, soils become less weatherable and the feedback's strength wanes, making a planet susceptible to catastrophic warming events or hard snowballs where the surface becomes frozen over. We find that tectonic uplift spurred by continental collisions cannot be the sole supplier of weatherable minerals within our model framework, as such climates either become uninhabitably hot (for complex life) as soils become leached of weatherable minerals or experience extreme swings in temperature over short timescales.  This conclusion is strengthened when taking into account the destabilizing effects of outgassing variability and increasing stellar luminosity.  In addition to frequent collisions, other resupply mechanisms for weatherable minerals, such as wind-driven dust transport, glacial erosion, and/or seafloor weathering, are likely required for long-term stability on Earth-like terrestrial exoplanets.

\end{abstract}

\section{Introduction}

Earth's surface temperature has remained `habitable'--permissive of surface liquid water--over almost all of its history \citep{Wilde2001,catling_kasting_2017}. The carbonate-silicate weathering feedback, with the first mathematical model proposed by \citet*{Walker1981} (hereon WHAK), has long been argued as the primary reason for Earth's climate self-regulation.  The bounds of where this feedback cycle can conceivably operate on terrestrial planets with \ce{N2-O2-H2O-CO2} atmospheres
have been used to define the `habitable zone' of extrasolar planets \citep{KASTING1993}. Without a fast-acting ($\sim100$ ky) negative feedback on changes in Earth's temperature, its habitability would quickly be jeopardized by small perturbations in incoming solar radiation (ISR) or volcanic outgassing of \ce{CO2} \citep{berner1997,zeebe2008close,Toby2020}. Whether the weathering feedback truly acts as Earth's thermostat or its impact on Earth's climate stability is over-emphasized due to survivorship bias remains an open question \citep{Toby2020,arnscheidt2022}. Thus, probing this feedback is crucial to understanding planetary habitability.

The feedback balances the release of carbon dioxide gas (\ce{CO2}) into the atmosphere through \ce{CO2} drawdown spurred by the dissolution of exposed silicate rocks in water (referred to as chemical weathering). Over time, these rocks become leached of minerals available for weathering, but the surface is resupplied through processes like tectonic uplift promoting erosion of the weathering zone.  Greenhouse gases (GHGs) such as \ce{CO2} are needed for liquid water at Earth's surface, as Earth's predicted surface temperature without the greenhouse effect is \SI{-18}{\celsius}. The atmospheric \ce{CO2} reservoir has a short ($\sim\SI{300}{kyr}$) residence time and can be perturbed by changes and imbalances in globally-integrated weathering or volcanic outgassing fluxes. These imbalances, in turn, directly affect the atmospheric $p$\ce{CO2} and can lead to climate warming or cooling.

The feedback also can be overwhelmed by external perturbations, such as increasing incoming solar radiation (ISR), that may lead to a runaway greenhouse effect where surface oceans evaporate completely, driving all life extinct \citep{goldblatt2012}, similar to what may have happened on Venus \citep{kasting1988}. Conversely, low ISR and/or GHG concentrations can lead to `snowball' events where the planet is entirely ice-covered \citep{Hoffman2017}. Little work has been done to quantify which planetary parameters are important for an Earth-like planet's stability despite a large variability in weathering spurred by plate collisions and changes in volcanic \ce{CO2} outgassing rates. Work on the stochastic randomness of volcanic outgassing rates of \ce{CO2} \citep{wordsworth21,baum2022snowball} and continental fraction/geometry \citep{baum2022b} over Earth's history in the context of initiating hard snowballs suggests that biological feedbacks to the silicate weathering cycle may have prevented snowball events throughout the Phanerozoic Eon ($\SI{541}{Mya}-$present).  However, these models assume that outgassing, continental geometry, and atmospheric \ce{CO2} concentrations behave as a purely random (stochastic) process, and do not track how weathering leaches minerals from soil. Many of these models also include a variety of processes that are likely unique to Earth, and lack general applicability to `Earth-like' planets.

Key open questions include: How does the land fraction of a planet affect its resilience?  How does the rate of volcanic outgassing of \ce{CO2} affect its long-term climate stability? Is resupply of minerals for weathering via tectonic uplift from collisions sufficient for long-term habitable climate?

In this work we present an idealized model, titled \texttt{DISKWORLD}, to test the resilience of the carbonate-silicate weathering feedback presented in \citet{Maher2014} and \citet{Winnick2018} against the randomness of continental plate collisions over geologic timescales for an Earth-like planet. In Section \ref{ch:2} we describe the carbonate-silicate weathering feedback including recent refinements. In Section \ref{ch:3} we  describe our model.  In Section \ref{ch:4} we present the results of our simulations over a wide variety of possible plate tectonic setups as well as test the effects of varying outgassing fluxes and our model's applicability to the faint young Sun paradox. In Section \ref{sec:disc} we discuss our results. In Section \ref{ch:conc} we present our conclusions and discuss future work regarding the role of tectonic collisions in climate stability.

\section{The Carbonate-Silicate Weathering Feedback} \label{ch:2}
Exposed silicates are dissolved via rainwater through reactions such as the following,

\begin{equation} \label{eq:cscycle1}
    \ce{CaSiO3(s) + H2O(l) +2 CO2(g) -> Ca^{2+}(aq) + 2HCO3^{-}(aq) + SiO2(aq)},
\end{equation}
where \ce{CaSiO3} is used to represent a complex combination of minerals for stoichiometric simplicity.  The products are washed to the ocean via rivers. If the ocean is saturated with respect to calcium carbonate (\ce{CaCO3}), which is generally the case for Earth \citep{Zeebe2012}, carbonate precipitates,

\begin{equation} \label{eq:cscycle2}
    \ce{Ca^{2+} + 2HCO3^{-} -> CaCO3 + H2O + CO2}.
\end{equation}
Combined, Reactions \ref{eq:cscycle1} and \ref{eq:cscycle2} remove one molecule of \ce{CO2} from the atmosphere in the form of \ce{CaCO3}. Some \ce{CaCO3} settles onto the seafloor and is subducted into Earth's mantle, and can be released as \ce{CO2} back into the atmosphere via metamorphic decarbonation and volcanism:

\begin{equation} \label{eq:cscycle3}
    \ce{CaCO3 + SiO2 -> CaSiO3 + CO2}.
\end{equation}

Thus the net effects of silicate weathering and volcanic outgassing can be summarized as:
\begin{equation} \label{rxn:cscyclenet}
    \ce{CaSiO3 + CO2  <=>[Weathering][Outgassing] CaCO3 + SiO2}.
\end{equation}

The balance between the fluxes in Reaction \ref{rxn:cscyclenet} sets the amount of inorganic carbon in the ocean plus atmosphere `slow carbon' system. Imbalances in these fluxes have been tied to some of Earth's most catastrophic extinction events as well as `Snowball Earth' episodes---periods of time when Earth's surface was mostly frozen \citep{Kirschvink1992,Hoffman1998,COX2016}. For example, the Paleocene-Eocene Thermal Maximum (PETM, $\sim55$ Mya) has been associated with a massive release of inorganic carbon into the atmosphere over a short ($20-50$ ky) timescale; up to $50\%$ of seafloor foraminifera went extinct \citep{Thomas1990,Wright2013,Bowen2014}.  In most cases, we do not know if these climate-destabilizing events were caused by changes in outgassing (source-driven) or weathering fluxes (sink-driven) \citep{Godderis2017}.

The rate of Reaction \ref{eq:cscycle1} is sensitive to local rainfall/evaporation, lithology, and the amount of fresh silicate rocks provided by tectonic uplift \citep{Graham2020,hakim2021}. Without tectonic activity or other processes resupplying fresh minerals, soil would become leached of reactants for the drawdown of \ce{CO2} in Reaction \ref{eq:cscycle1}, and atmospheric \ce{CO2} levels would quickly become large enough to cause acute global warming and mass extinction.

\begin{figure}
    \centering
    \includegraphics[width=0.9\linewidth]{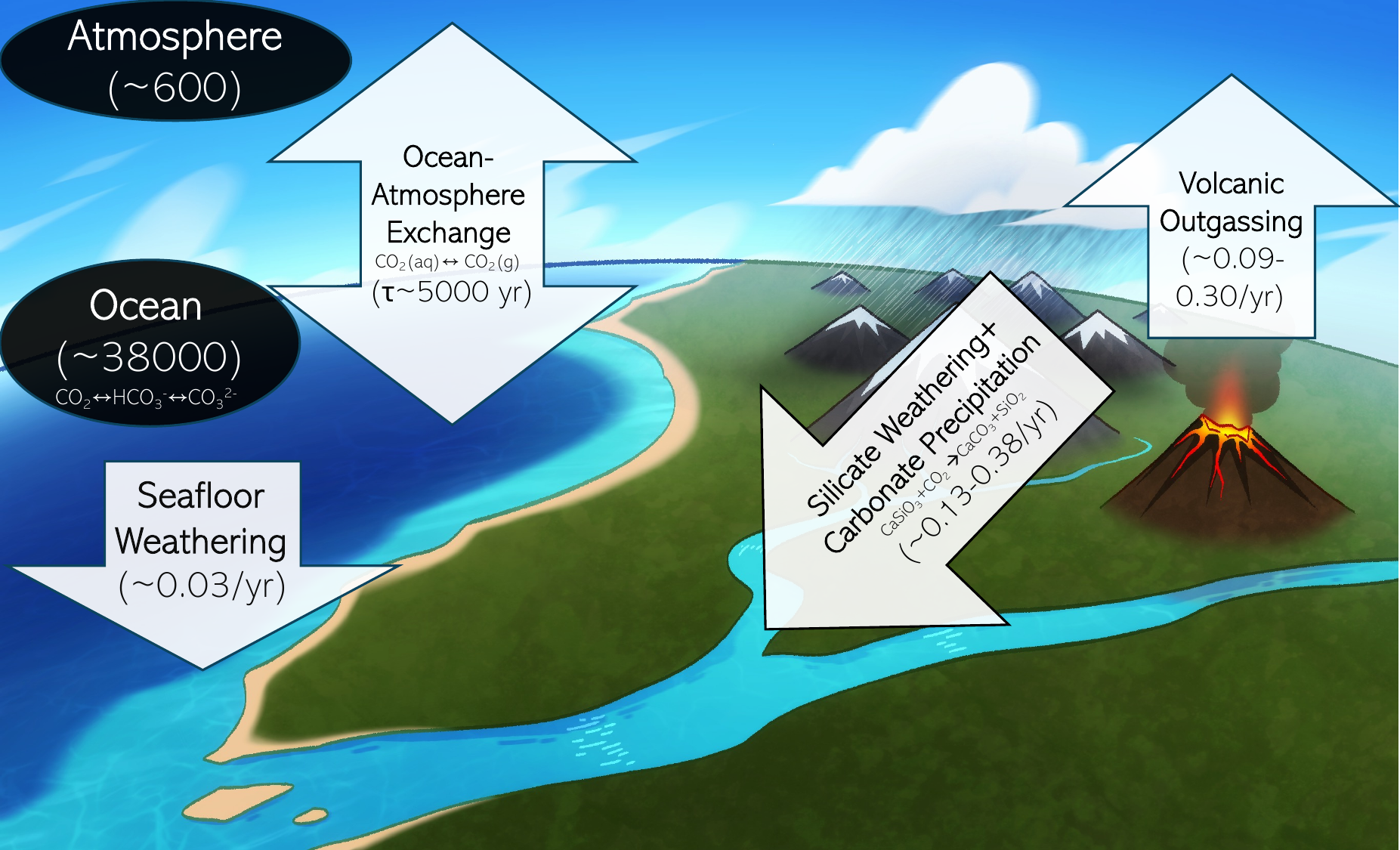}
    \caption{Modern-day Earth estimates of reservoirs and fluxes of the slow carbon system considered in our model, in units of gigatonnes of carbon (GtC, $10^{12}~\si{kg C}$) from \citet{Lee2019}.  There is significant uncertainty in modern-day flux values. Carbon in continental crust ($\sim\SI{42e6}{GtC}$), continental lithospheric mantle ($\lesssim\SI{48e6}{GtC}$), and oceanic crust/lithospheric mantle ($\sim\SI{14e6}{GtC}$) are not tracked in our simulations, as they are much larger than surface reservoirs and are treated as functionally inexhaustible.}
    \label{fig:creservoir}
\end{figure}

Physical erosion of hills and mountains that form when continents collide exhumes underlying bedrock that is rich in the cations needed for Reaction \ref{eq:cscycle1}. When a large amount of weatherable minerals is introduced through these collisions, low \ce{CO2} concentrations may cause most of Earth to freeze over. Figure \ref{fig:creservoir} shows a simplified overview of the reservoirs and fluxes of carbon considered in this work.

\subsection{The WHAK Model}
\citet*{Walker1981} propose a model for global weathering from river catchments and suggest a negative feedback capable of keeping Earth's climate stable on geologic timescales. In this model, the rate of silicate rock weathering, $W$ (in \si{t.km^{-2}.yr^{-1}}), is set by the globally averaged temperature $T$, river runoff rate $q$ (roughly proportional to the precipitation rate), and partial pressure of \ce{CO2} ($p$\ce{CO2}):

\begin{equation}\label{eq:WHAK}
    W= kqe^{\frac{T-T_{ref}}{17.7\si{K}}}\left(\frac{p\ce{CO2}}{p\textrm{CO$_{2,\textrm{\,ref}}$}}\right)^{0.3}.
\end{equation}
Reference values for $T$ and $p$\ce{CO2} are typically set at \SI{12}{\celsius} and 280 ppm, respectively.  Note that the e-folding temperature becomes \SI{13.7}{K} if assuming the exponential relation of $q\propto e^{(T-T_{ref})/\SI{60}{K}}$ used in \citet{Walker1981}, but \SI{17.7}{K} when including it as a separate parameter. This approach has been used in numerous studies, including reconstructions of Earth's past \ce{CO2} inventory \citep[e.g.,][] {Godderis2017,Godderis2017b,Krissansen2017}  and testing the model's sensitivity to parameters such as land fraction, position, and variable volcanic outgassing (e.g. \citealt{Graham2020,baum2022snowball,baum2022b}). However, the WHAK model does not track how soils evolve as they are leached of weatherable minerals; the weatherability factor $k$ lacks a physical interpretation.

Equation \ref{eq:WHAK} contains a negative (stabilizing) feedback---if $p$\ce{CO2} or temperature increases, weathering will increase, increasing \ce{CO2} drawdown and thereby rebalancing weathering and outgassing fluxes.  In addition, runoff $q$ is expected to increase or decrease alongside temperature, further cementing the negative feedback.  This model gives a roughly constant \textit{climate feedback strength} $\alpha$ \citep[referred to as `climate sensitivity' in][]{Maher2014}, which we define as the percentage change in weathering per increase in temperature, of $\sim\SI{15}{\%.\celsius^{-1}}$.  Higher values of $\alpha$ imply a stronger (more stabilizing) negative feedback.

\subsection{The MAC Model} \label{sec:MAC}

\citet{Maher2014} (hereon MAC) present a solute transport model of the carbonate-silicate weathering feedback. Their analysis also uses weathering flux and runoff data for modern rivers. In their model, the weathering flux of dissolved minerals to the ocean is related to the ratio of the mean fluid travel time through a reactive assemblage ($T_{f}\approx L\phi/q$) to the time required for the fluid to reach chemical equilibrium ($T_{eq}\approx C_{eq}/R_{n}$). As $T_{f}/T_{eq}$ increases, solute concentration in the river asymptotes to a thermodynamic maximum concentration $C_{eq}$. The ratio of the two timescales, $T_{f}/T_{eq}$, is known as the `Damk$\ddot{\textrm{o}}$hler number' \citep{boucher1963}. Here, $L$ is the reactive path length that rainwater travels through, $\phi$ is the porosity of the soil, and $R_{n}$ is the reaction rate. Most cation release from weathering occurs in this upper layer of soil defined by $L$ and not in the underlying bedrock \citep{Brantley2023}.

Continuing to follow \citet{Maher2014}, $R_{n}=\rho_{sf}k_{eff}AX_{r}f_{w}$, where $\rho_{sf}$ is the ratio of solid mineral mass to fluid volume (defined as $\rho_{sf}=\rho/\phi$, where $\rho$ is the soil density), $A$ is the specific area for the minerals in question, $X_{r}$ is the fraction of reactive material in fresh (unweathered) rock, $k_{eff}$ is the weathering rate constant, and $f_{w}$ is the fraction of fresh rock in the assemblage.  $f_{w}=(1+mk_{eff}A\tau_{s})^{-1}$ where $\tau_{s}$ is the effective soil age (or soil residence time) and $m$ is the molar mass of the weatherable minerals.  Here we use $\tau_{s}$ instead of the original $T_{s}$ to avoid confusion with temperature $T$. Soil age is related to the erosion rate $E$ (in \si{m.yr^{-1}}) as $E=h/\tau_{s}$, where $h$ is the erosion zone thickness, and is meant to be a proxy for the amount of fresh rock available for weathering.  \citet{Maher2014} fit for $R_{n,max}$ ($R_{n}$ value when the assemblage is all fresh minerals, $f_{w}=1$) through the reactive transport model CrunchFlow \citep{steefel2009} using an idealized reaction meant to represent silicate weathering on a global scale, the dissolution of \ce{An20} plagiocase to halloysite/kaolinite:

\begin{equation}\label{rxn:plag}
\ce{1.67Ca_{0.2}Na_{0.8}Al_{1.2}Si_{2.8}O8 + 2CO2(g) + 3H2O <-> Al2Si2O5(OH)4 + 0.33Ca^{2+} + 1.33Na+ + 2.67SiO2(aq) + 2HCO3-}.
\end{equation}

\citet{Maher2014} introduce a quantity $Dw$ to quantify the efficiency of solute production,
\begin{equation} \label{eq:dw}
    Dw=\frac{L\phi}{T_{eq}}=\frac{L\phi R_{n,max}}{C_{eq}(1+mk_{eff}A\tau_{s})}
\end{equation}
(note that $Dw$ is a single variable). Like the constant $k$ in the WHAK model, $Dw$ varies with climate and lithology, from tectonically inactive regions (cratons, $Dw\sim\SI{0.01}{m.yr^{-1}}$) to tectonically active mountain ranges ($Dw\sim\SI{0.3}{m.yr^{-1}}$). \citet{Maher2014} use this coefficient to calculate the weathering flux of aqueous silica (\ce{SiO2}), $W$ (in \si{t.km^{-2}.yr^{-1}}), where
\begin{equation} \label{eq:MACw}
W=Cq=C_{eq}\frac{e^{2}Dw}{1+e^{2}Dw/q},
\end{equation}
and $e$ is Euler's constant. While MAC use the terms `mountainous' and `cratonic' for high and low values of $Dw$, respectively, these are more accurately thought of as `tectonically active/high erosion' and `tectonically inactive/low erosion' zones.  Tectonically inactive mountain ranges, such as the Appalachians, would have `cratonic' properties, due to relaxed present-day erosion rates compared to when they formed \citep{matmon2003}.

\begin{figure}
    \centering
    \includegraphics[width=.8\linewidth]{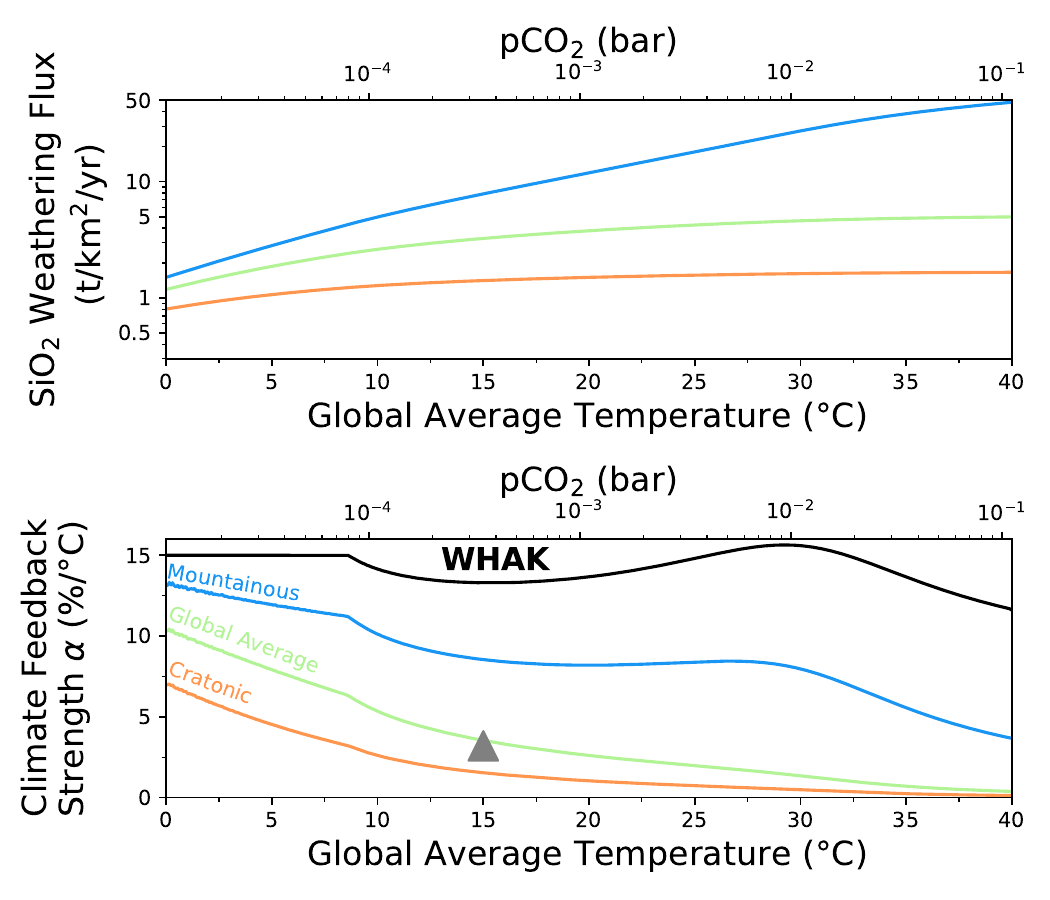}
    \caption{
    (Top) \ce{SiO2} weathering flux as a function of globally averaged surface temperature for various soil ages determined by the `mountainous' (tectonically active/high erosion) $Dw=\SI{0.3}{m.yr^{-1}}$ (blue), `global average' \SI{0.03}{m.yr^{-1}} (green), and `cratonic' (tectonically inactive/low erosion) \SI{0.01}{m.yr^{-1}} (orange) values from \citet{Maher2014} Fluxes are calculated assuming a latitude of $\Phi=0\degree$; most weathering occurs at low latitude in our model. 
    (Bottom) Climate feedback strength $\alpha$ as a function of global temperature. A higher value of $\alpha$ represents a climate more stable to perturbations in weathering or outgassing fluxes. MAC model values are presented alongside the WHAK formulation (Equation \ref{eq:WHAK}, black). Wiggles in feedback strength are due to the non-linear dependence of temperature as a function of $p$\ce{CO2} in our model. 
    The gray triangle represents an estimate for the modern-day globally averaged value of $\alpha$ from \citet{Brantley2023}, which is very close to the value calculated for the global average $Dw$.
    }
    \label{fig:feedback}
\end{figure}

In the high $Dw$ limit (enough time to reach equilibrium concentrations), $W\approx C_{eq}q$ and weathering becomes \textit{transport limited} by river runoff as predicted by Equation \ref{eq:WHAK}. Conversely, in the low $Dw$ limit, $W\approx C_{eq}e^{2}Dw$, and weathering becomes \textit{kinetically limited} by $Dw$, asymptotically approaching $C_{eq}e^{2}Dw$ as $q\longrightarrow\infty$.  These limits can be seen in high and low $Dw$ value soils in Figure \ref{fig:feedback}. Thus, the negative stabilizing feedback of the MAC model relies on the assumption that global precipitation will increase with increasing temperatures (discussed in Section \ref{sec:precip}). Due to this assumption ($q$ increasing with increasing temperatures), higher values of $Dw$ increase the system's resilience to changes in temperature, whereas low values of $Dw$ are relatively insensitive to changes in temperature. 

Although temperature does affect the kinematics of the dissolution reactions and thus $k_{eff}$,
\begin{equation}
\label{eq:MACtemp}
    \frac{k_{eff}(T)}{k_{eff}(T_{ref})}=e^{\frac{E_{a}}{R}\left(\frac{1}{T_{ref}}-\frac{1}{T}\right)},
\end{equation}
where $E_{a}$ is the apparent activation energy of mineral dissolution and $R$ is the gas constant, \citet{Maher2014} claim that these effects are small and that the weathering rate is primarily set by soil properties and local runoff (these assumptions are explored in Appendix \ref{sec:tdepend}). We find that temperature dependencies have a significant influence on weathering fluxes and thus include them in our model.
We also include additions from \citet{Winnick2018}, who find that $C_{eq}$ is sensitive to local $p$\ce{CO2},
\begin{equation}\label{eq:pco2dep}
    C_{eq}(p\ce{CO2})=C_{eq}(p\textrm{CO$_{2,\textrm{\,ref}}$})\left(\frac{p\ce{CO2}}{p\textrm{CO$_{2,\textrm{\,ref}}$}}\right)^{0.316}.
\end{equation}
While some previous studies (e.g., \citealt{baum2022snowball}) have neglected the $p$\ce{CO2}-dependence from \citet{Winnick2018}, we find that its inclusion has a drastic effect on climate feedback strength.  We discuss the formulation and implications of this effect in Appendix  \ref{sec:tdepend}.

A key advantage of the MAC model is that we are able to track the time-evolution of weathering fluxes in aging soils. Previous implementations of the WHAK model assume that weathering is primarily constrained by local climate types and not soil properties. For example, \citet{Godderis2017b} use local temperature and runoff to define six climate types that in turn determine $k$ used in Equation \ref{eq:WHAK}. However, $k$ should decrease over time as the local rock is leached of cations assuming there is no background resupply mechanism, so climate alone cannot control the spatial variation of weathering flux \citep{west2005}. Varying soil age allows us to simulate the effect of resurfacing that restores minerals for the carbonate-silicate weathering cycle.  For these reasons, our model largely follows MAC instead of WHAK.  We note that, as found in previous studies \citep{baum2022b},  MAC underestimates weathering fluxes when applied on a global scale on the order of $\sim3-4\times$ lower than modern-day estimates of \ce{CO2} outgassing/silicate weathering fluxes from \citet{Lee2019}. There is some tension in recent estimations of global \ce{CO2} weathering and outgassing fluxes, with most studies adopting a rate of $F_{vol}\approx\SI{7.5}{Tmol.yr^{-1}}$ (e.g., \citealt{Krissansen2018,baum2022snowball,baum2022b}), whereas some studies have estimated considerably lower outgassing rates ($\sim\SI{3.4}{Tmol.yr^{-1}}$, \citealt{coogan20}), more in-line with our model's output. In contrast to \citet{baum2022b}, we do not alter any of the original MAC formulation parameters as it would affect the model's fundamental predicted feedback strength. This choice reduces \ce{CO2} drawdown, and may lead to an overabundance of warming events and under-abundance of cooling events. The effects of increasing MAC weathering fluxes to match those of modern-day Earth are discussed in Appendix \ref{sec:ap_fudge}.

Recent studies have constrained Earth's globally-averaged climate feedback strength. Synthesizing laboratory dissolution data over various lithologies and accounting for the distribution of kinetic-, erosive-, and runoff-limited regions on Earth, \citet{Brantley2023} estimated a present-day global average value of $\alpha$ to be \SI{3.2}{\%.\celsius^{-1}}. Using the global average value of $Dw=\SI{0.03}{m.yr^{-1}}$, we find a similar value of $\alpha=\SI{3.5}{\%.\celsius^{-1}}$.

\section{Methods} \label{ch:3}

\begin{deluxetable*}{llll}
\tablecaption{\label{tab:parameters} Definition of select model parameters.  For a more detailed table of weathering flux-related parameters, see the supplemental material of \citet{Maher2014}.}
\tablewidth{0pt}
\tablehead{
\colhead{Parameter} & \colhead{Definition} & \colhead{Units} & \colhead{Value}}

\startdata
$R_{pl}$ & Planetary radius & m & \SI{6.371e6}{}\\
$S_{0}$ & Solar insolation flux & \si{W.\metre^{-2}} & 1361\\
$A$ & Planetary albedo & & 0.3 \\
\hline
\ce{pCO2_{,0}} & Initial atmospheric \ce{CO2} concentration & \si{\micro bar} & 280\\
$T_{0}$ & Initial globally-averaged surface temperature & \si{\celsius} & 13 \\
$C_{ocean}$ & Initial ocean inorganic carbon reservoir & GtC & 41000 \\
$V_{ocean}$ & Ocean volume & L & Variable, default $1.33\times10^{21}$ \citep{charette2010}\\
$S$ & Ocean salinity & g/L & 35\\
$\overline{p}_{ref}$ & Modern global average precipitation & \si{m.yr^{-1}} & 1.0 \citep{Xie1997} \\
$\Gamma$ & Runoff coefficient & & 0.259 \citep{ghiggi2019}\\
\hline
$L$ & Flow path length & m & 1 \\
$\phi$ & Soil porosity &  & 0.175 \\
$C_{eq,\textrm{\,ref}}$ & Reference equilibrium solute concentration & \si{\micro M} & 375 \\ 
$R_{n,max}$ & Maximum reaction rate & \si{\micro M.yr^{-1}} & 1085 \\
$k_{eff}$ & Reference rate constant & \si{mol.m^{-2}.yr^{-1}} & $8.7\times10^{-6}$\\
$mAk_{eff}$ & $f_{w}$ Scale factor for \ce{SiO2} & yr$^{-1}$ & $8.5\times10^{-5}$ \\
$E_{a}$ & Silicate weathering activation energy & \si{kJ.mol^{-1}} & 38 \\
$\tau_{s,mountain}$ & Soil age during a collision & yr & \SI{8e3}{} \\
$\tau_{s,max}$ & Maximum effective soil age & yr & \SI{6e5}{} \\
\hline
$R_{disk}$ & Disk radius & m & Variable, default \SI{2.43e7}{}\\
$N_{disk}$ & Number of disks &  & Variable, default 8 \\
$V$ & Average disk velocity & \si{cm.yr^{-1}} & 4 \\
$\sigma_{V}$ & Disk velocity standard deviation & \si{cm.yr^{-1}} & 2 \\
\hline\hline
$F_{vol,\bigoplus}$ & Modern \ce{CO2} outgassing flux & \si{GtC.yr^{-1}} & $\sim0.085-0.300$ \citep{Lee2019} \\
$W_{\mathrm{CO_2},\oplus}$ & Modern silicate weathering \ce{CO2} drawdown & \si{GtC.yr^{-1}} & $\sim0.13-0.38$ \citep{Lee2019} \\
$W_{SFW,\mathrm{CO_2}}$ & Modern seafloor weathering \ce{CO2} drawdown & \si{GtC.yr^{-1}} &  0.0054 \citep{coogan2013} \\
\enddata
\end{deluxetable*}

Our goal is to test how a world with an initial balance between volcanic outgassing and weathering sequestration of \ce{CO2} responds to long-term evolution and short-term perturbations in volcanic outgassing and in weathering fluxes due to tectonic drift. The model is applicable to planets with Earth-like land area and those with different plate configurations. Default parameters can be seen in Table \ref{tab:parameters}.

\subsection{Disks}\label{sec:disks}
We model an Earth-sized planet (radius $R_{pl}=R_{\bigoplus}$). Disks are defined as spherical caps with radius $R_{disk}$, velocities that are randomized with mean speed $V_{disk}$ and standard deviation $\sigma_{V}$, and randomized starting locations.  We use Euler poles to describe disk motion, where disk centers follow a great circle around a rotational Euler pole (a randomly chosen longitude and latitude) chosen for each individual disk (see Appendix \ref{sec:ap_euler}).  Collisions between disks do not alter their shapes and velocities remain constant over time. This conserves total disk volume while the total land surface area is allowed to evolve over time as disks slide under/over each other. Although it is possible that continental plates have slowed as Earth's interior has cooled over Gyr timescales, this process is not well understood, and some studies have suggested that plate speeds were \textit{slower} in the past (e.g., \citealt{korenaga2006}). We prescribe $V_{disk}=\SI{4}{cm.yr^{-1}}$ and $\sigma_{V}=\SI{2}{cm.yr^{-1}}$, in-line with modern-day estimates of continental plate speeds from \citet{zahirovic2015}.  

For book-keeping, disks are sub-divided into equiareally spaced points that co-move with the disk's center. Points are considered `colliding' when they lie within the boundary of another disk.  Collisions reset the effective soil age of the point to a value $\tau_{s,mountain}$, corresponding to the uplift of new soil and creation of mountain belts caused by plate collisions. 
Points that are not colliding `age' over time---their effective soil ages increase and thus erosion rates decrease.  This represents local soils becoming depleted of cations available for weathering over time. 
$\tau_{s,mountain}$ was determined by fitting $\tau_{s}$ in Equation \ref{eq:dw} to the $Dw$ value defined in \citet{Maher2014} for `mountainous' regions, \SI{0.3}{m.yr^{-1}}.  This gives $\tau_{s,moutain}\approx8000$ yr. $\tau_{s}$ values, alongside approximate erosion rates, for a variety of $Dw$ regimes can be seen in Table \ref{tab:dwage}.

\begin{deluxetable*}{llllr}
\tablecaption{\label{tab:dwage} Soil $Dw$ values and associated soil ages for different regimes presented in \citet{Maher2014}.  When soils are young, fresh weatherable materials are more abundant. All soil ages are calculated using default parameters presented in Section \ref{sec:MAC} and Table \ref{tab:parameters}. Erosion rate calculations assume the flow path length and the erosion/weathering zone thickness are equal (1 m). 
}
\tablewidth{0pt}
\tablehead{
\colhead{$Dw$ (\si{m.yr^{-1}})} & \colhead{$\tau_{s}$ (yr)} & \colhead{Effective Erosion Rate (\si{\micro\metre.yr^{-1}})} & \colhead{Fraction of Fresh Rock $f_{w}$} & \colhead{Notes}}
\startdata
0.3 & \SI{8e3}{} & 120 & 0.59 & Mountainous \\
0.03 & \SI{2e5}{} & 5 & 0.058 & Global Average \\
0.01 & \SI{6e5}{} & 2 & 0.019 & Cratonic \\
0.003 & \SI{2e6}{} & 0.5 & 0.0058 & Global Minimum $Dw$\\
\enddata
\end{deluxetable*}

\subsection{$p$\ce{CO2} and Global Temperatures} \label{sec:pco2t}

In order to estimate globally-averaged temperature $\overline{T}$ as a function of $p$\ce{CO2}, we use the polynomial fits derived in \citet{Krissansen2018} from the output of their 1D radiative-convective model.  Their results are fit for $10^{-4}~\si{bar}<p$\ce{CO2}$<10~\si{bar}$ and $0.7L_{\odot}<L<L_{\odot}$, where $L_{\odot}$ is modern-day solar luminosity.  We include additions from \citet{Kadoya2020}, who assumed a log-linear relationship below $p$\ce{CO2}$=10^{-4}~\si{bar}$, to estimate temperatures for low ($10^{-6}-10^{-4}$ bar) $p$\ce{CO2}.  We note that this approach neglects the effects of ice-albedo feedback (albedo is set at 0.32), and may overestimate temperatures at low ($<\SI{1e-4}{bar}$) $p$\ce{CO2}. Assuming a pre-industrial $p$\ce{CO2} of $\SI{280}{\micro bar}$, this leads to a corresponding pre-industrial global temperature of \SI{13}{\celsius}.

\subsection{Precipitation and Runoff} \label{sec:precip}

\begin{figure}
    \centering
    \includegraphics[width=.5\linewidth]{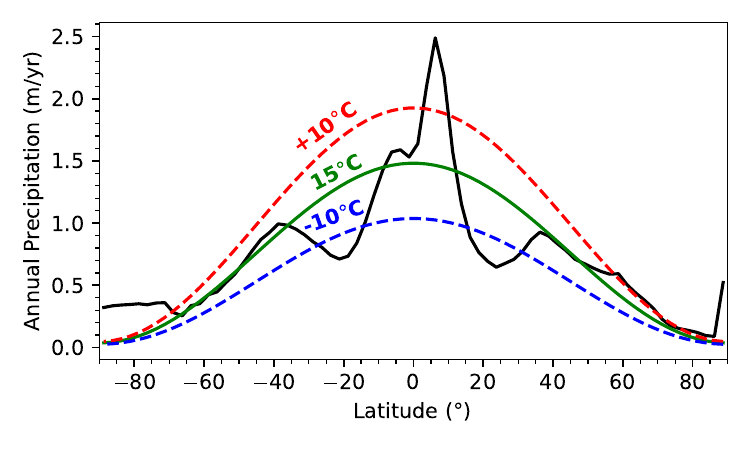}
    \caption{A comparison of modern-day Earth precipitation estimates \citep[black, from the Climate Prediction Center Merged Analysis of Precipitation,][]{Xie1997} to our simplified latitudinal dependence of precipitation (green, Equation \ref{eq:latprecip}). Red and blue dotted lines indicate effects of a \SI{10}{\celsius} warming and cooling, respectively.}
    \label{fig:latprecip}
\end{figure}

The MAC model creates a climate-stabilizing negative feedback if rainfall and thus river runoff increase as a planet's surface temperature increases. This happens due to precipitation balancing increasing evaporation rates. Following \citet{Graham2020}, we linearize our globally averaged precipitation's ($\overline{p}$) temperature dependence:

\begin{equation} \label{eq:precip}
    \overline{p}(\overline{T})=\overline{p}_{ref}(1+\epsilon(\overline{T}-\overline{T}_{ref})),
\end{equation}
where $\overline{T}$ is the globally averaged surface air temperature and $\epsilon$ is the fractional change in precipitation per change in temperature measured at a reference temperature and precipitation value.  $\overline{p}_{ref}$ and $\overline{T}_{ref}$ correspond to modern-day Earth values [$\overline{p}_{ref}=\SI{1}{m.yr^{-1}}$ \citep{Xie1997} and $\overline{T}_{ref}=15\degree$C]. We set $\epsilon=\SI{3}{\%.\celsius^{-1}}$ following \citet{Graham2020}. In reality, $\epsilon$ likely varies significantly with latitude \citep{Manabe2004,OGorman2009,ogorman2015}.  Zonally averaged temperature (per latitude $\Phi$) is solved for using the simplified treatment in \citet{wang1980}, i.e. 

\begin{equation}\label{eq:tlat}
    T(\Phi)=\overline{T}-\SI{32.1}{\celsius}\times\frac{3\sin^{2}\Phi-1}{2},
\end{equation}
leading to a surface temperature of \SI{29}{\celsius} at the equator and and \SI{-19}{\celsius} at the poles for the pre-industrial value of $p$\ce{CO2}. 
Zonally-averaged precipitation is then solved for via
\begin{equation}\label{eq:latprecip}
    p(\overline{T},\Phi)=\overline{p}(1+\epsilon(T(\Phi)-\overline{T}))=\overline{p}(1+\epsilon(\SI{-32.1}{\celsius}\times\frac{3\sin^{2}\Phi-1}{2})).
\end{equation}
Our simplified latitudinal dependence is compared to values for modern-day Earth \citep{Xie1997} in Figure \ref{fig:latprecip}. Recent studies have called into question the assumption that precipitation increases with increasing global temperatures (e.g., \citealt{pierrehumbert2002,Ogorman2008,le2009}), and climate models have shown that increasing temperatures may cause precipitation to \textit{decrease} when hotter than $\sim47-\SI{57}{\celsius}$ \citep{liu24}. In this study, we consider climates cooler than \SI{40}{\celsius}, which is within the semi-linear $p-T$ relationships predicted by climate models in \citet{liu24}.  However, this effect needs to be taken into consideration when analyzing hothouse climates.

Again following \citet{Graham2020}, we take river runoff $q$ to be a linear function of precipitation:

\begin{equation} \label{eq:runoff}
    q=\Gamma p.
\end{equation} $(1-\Gamma)$ represents the fraction of precipitation lost to evaporation before being converted to runoff. We set $\Gamma=0.259$ based on globally-integrated runoff estimates from \citet{ghiggi2019}.

Thus, the planet-integrated \ce{SiO2} weathering flux is
\begin{equation} \label{eq:totalw}
   W_{\ce{SiO2}}(t)=\frac{A_{land}(t)}{n_{grid}}\sum_{i=1}^{n_{grid}} W_{i}(\tau_{s,i},k_{eff}(T_{i}(p\ce{CO2}),\Phi_{i}),
\end{equation}
where $A_{land}$ is the total exposed land area and $n_{grid}$ is the total number of grid points. $A_{land}$ varies with $t$ because collisions reduce the total amount of exposed land area. $\tau_{s}$, $T$, and $\Phi$ are also functions of $t$ for each grid point. \textit{Weathering flux}, described by Equation \ref{eq:totalw} differs from \textit{weatherability}, which we define as the weathering flux corresponding to a configuration of disks at \textit{given} values of $p$\ce{CO2} and $T$ (\SI{280}{\micro bar} and \SI{13}{\celsius}, respectively). \textit{Weatherability} does not take into account the time evolution of the climate given imbalances in weathering and outgassing. A planet with a larger fraction of colliding area may have a lower weathering flux at low temperatures than one with a low fraction of colliding area at high temperatures, but will have a higher weatherability.

\subsection{\ce{SiO2} fluxes and \ce{CO2} Consumption}
In order to relate the \ce{SiO2} fluxes calculated in Equation \ref{eq:totalw} to \ce{CO2} drawdown, we interpolate the results of \citet{GAILLARDET1999}, who compiled the concentrations of dissolved solids (including \ce{Ca^{2+}}, \ce{Mg^{2+}}, \ce{HCO3-}, and \ce{SiO2}) and discharge rates for 60 large rivers. They also estimated the atmospheric \ce{CO2} consumption from silicate weathering for each river.  We exclude 11 rivers that are presented as highly-polluted. By weighting the \ce{CO2}/\ce{SiO2} ratio of each river by its discharge of dissolved solids, we find an average molar \ce{CO2}/\ce{SiO2} ratio of 1.087---for every mole of \ce{SiO2} weathered and flushed out to the ocean, 1.087 moles of \ce{CO2} are removed from the atmosphere (Figure \ref{fig:co2sio2}). Thus, the \ce{CO2} drawdown (in \si{mol.\ce{CO2}.yr^{-1}}) is
\begin{equation}
    W_{\ce{CO2}}(t)=1.087\, W_{\ce{SiO2}}(t).
\end{equation}

\subsection{Seafloor Weathering}

Another possible source of negative feedback in the global carbon cycle is seafloor weathering (SFW). SFW occurs when seawater reacts with oceanic crust basalt to release soluble cations. Cations can then react with bicarbonate ion via Reaction \ref{eq:cscycle2} to remove inorganic carbon from the ocean-atmosphere system. Carbon drawdown from SFW on modern Earth is estimated as $\sim8-23\%$ of that of silicate weathering \citep{Lee2019}.  Evidence suggests that this proportion could have been much higher in the Archean Eon, possibly being the dominant \ce{CO2} sink \citep{Nakamura,Krissansen2018}.

The overall temperature, $p$\ce{CO2}, and ocean \ce{pH} dependencies of seafloor weathering are not well-constrained. Due to this, we treat seafloor weathering as a constant source of weathering only dependent on global ocean fraction, i.e.,
\begin{equation}
    W_{SFW}(\gamma)=W_{SFW,\oplus}\frac{(1-\gamma)}{0.7},
\end{equation}
where $W_{SFW,\oplus}=\SI{0.0054}{GtC.yr^{-1}}=\SI{0.45}{TmolC.yr^{-1}}$ is calibrated to modern-day estimates for \ce{CO2} drawdown due to  seafloor weathering \citep{coogan2013,Krissansen2018} and $\gamma$ is the surface area land fraction.  We explore including more complex relationships between seafloor weathering, ocean pH, and temperature at the pore spaces where weathering occurs in Section \ref{sec:sfw}. Including SFW as an effective `baseline' weathering flux  prevents trials with low amounts of land-area/initial collisions from having unrealistically low initial outgassing fluxes.

\subsection{Outgassing and Time Evolution}

The model calculates the volcanic outgassing flux of \ce{CO2}, $F_{vol}$, by assuming that outgassing and \ce{CO2} drawdown from silicate weathering are initially in balance:
\begin{equation}
    F_{vol}=1.087\, W_{\ce{SiO2}}(t=0)+W_{SFW}.
\end{equation}

This leads to different outgassing fluxes for different continental configurations, with higher $F_{vol}$ for setups with initially colliding disks. $F_{vol}$ remains constant throughout each simulation run (changed in Section \ref{sec:fvolpert}). While $F_{vol}$ has likely changed significantly over Earth's 4 Gyr history (e.g., \citealt{Krissansen2017}), the exact long-term behavior of \ce{CO2} outgassing is not well-constrained, and we aim to isolate the effects of tectonic noise on long-term habitability.

\begin{figure}
\centering
\includegraphics[width=\linewidth]{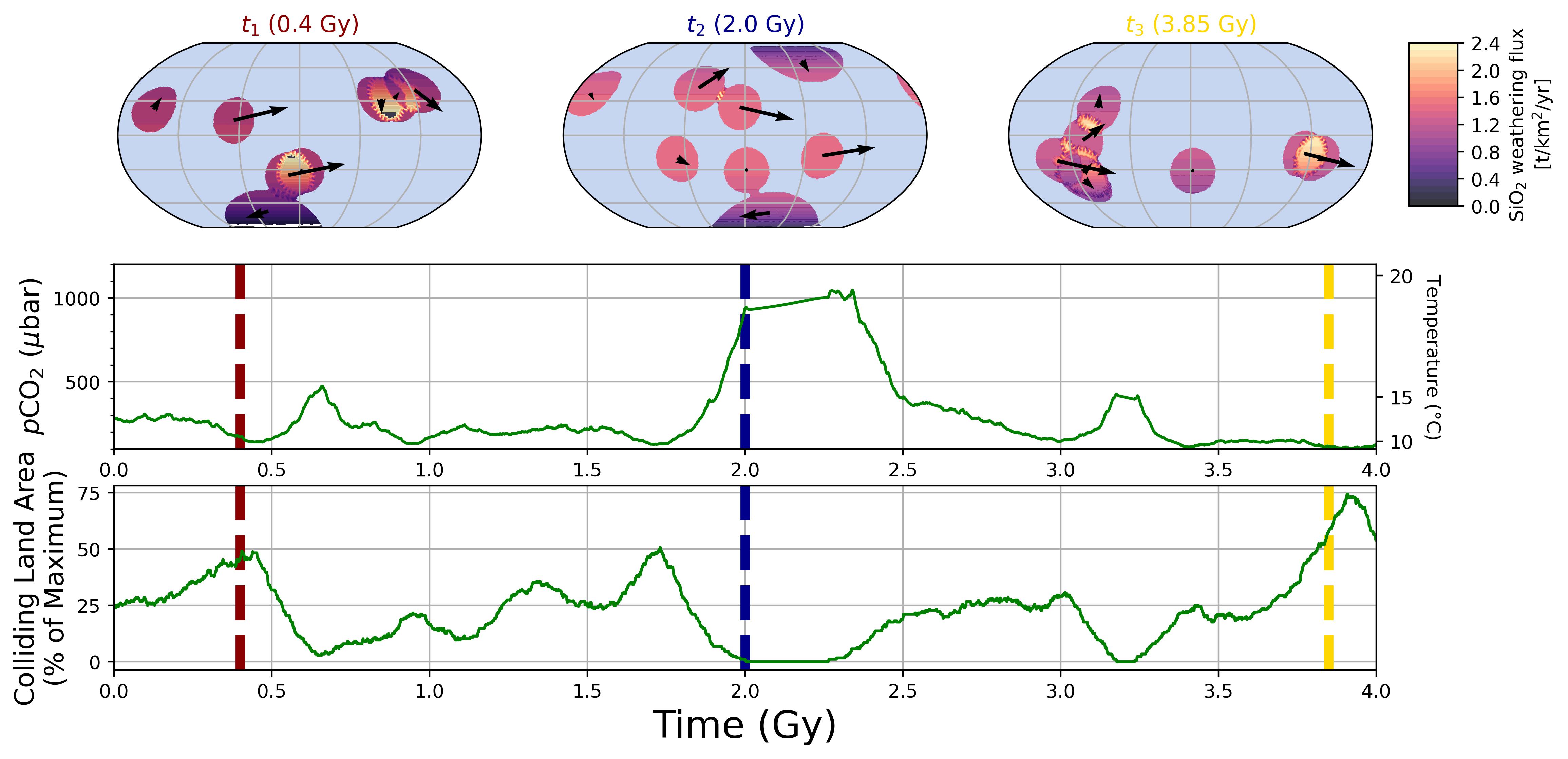}
\caption{An example run of the base \texttt{DISKWORLD} model spanning 4 Gy, showing disk configurations and localized weathering fluxes at three select times.  Arrows represent disk velocities.  $t_{1}$ shows how the introduction of colliding disks leads to a gradual decrease in atmospheric $p$\ce{CO2} and global temperatures.  At $t_{2}$, although there are no currently-colliding disks, the high $p$\ce{CO2} and global temperatures lead to increased global weathering fluxes owing to the carbonate-silicate weathering feedback (see Figure \ref{fig:feedback}), which is strong enough even in `cratonic' soils to prevent massive \ce{CO2} buildup.  $t_{3}$ represents an example state with large amounts of colliding disks near the equator, greatly increasing global weathering fluxes, potentially leading to a future snowball event.}
\label{fig:snapshot}
\end{figure}

Initial $p$\ce{CO2} and temperature conditions are chosen to mimic that of pre-industrial Earth [$p\ce{CO2}(t=0)=\SI{280}{\micro bar}$, $\overline{T}(t=0)=\SI{13}{\celsius}$]. Zonal temperatures are calculated as a function of $p$\ce{CO2} (Section \ref{sec:pco2t}). Disk locations, global weathering fluxes, ocean and atmospheric carbon reservoir content, temperature, and ocean pH are all calculated in adjustable time steps. We use an adaptive time step $dt\gg1000$ years so that the ocean and atmosphere equilibrate. Changes in the total inorganic carbon reservoir are calculated per time step as 

\begin{equation}
    \frac{\Delta C(t)}{dt}=W_{\ce{CO2}}(t)-F_{vol}.
\end{equation}
We then calculate the effects of $\Delta C$ on ocean-atmosphere partitioning (see Appendix \ref{sec:csys} for details). At each time step we record globally-integrated \ce{CO2} drawdown weathering flux, atmospheric $p$\ce{CO2}, ocean dissolved inorganic carbon content, and globally averaged surface temperature $T$. An example run of the \texttt{DISKWORLD} model is shown in Figure \ref{fig:snapshot}.

\subsection{Climate Catastrophes}

In order to study the climate stability of a given setup, we monitor for two possible `catastrophes'---uninhabitably hot `catastrophic warming' events and hard snowball states.

\subsubsection{Catastrophic Warming}
The effect of adding large amounts of \ce{CO2} to the atmosphere on surface temperatures has been studied in detail. While the runaway greenhouse---where greenhouse gases effectively prevent the planet from cooling until reaching extremely high surface temperatures---is likely unreachable at present-day or lower solar luminosity \citep{Goldblatt2013}, several studies suggest that a `moist greenhouse' state, where water vapor is able to bypass the atmospheric cold trap and reach high concentrations in the stratosphere, may occur for $p$\ce{CO2} as low as 6 mbar \citep{Ramirez2014} or 10 mbar \citep{kasting1986} ($\sim27-\SI{29}{\celsius}$ globally-averaged temperature in our model). This moist state can lead to the gradual loss of a planet's ocean due to enhanced photolysis of water vapor in the upper atmosphere followed by hydrogen escape to space. However, the $p$\ce{CO2} threshold for entering a moist greenhouse is still debated (e.g., \citealt{wolf15}), and atmospheric loss would need to be modeled in detail to determine ocean-loss timescales.

Instead of focusing on these two `greenhouse' scenarios, we consider an uninhabitably hot average surface temperature of \SI{40}{\celsius}, corresponding to the life cycle limits for most eukaryotes \citep{clarke_2014}.  While this would not necessarily spell the end of all life on Earth (bacteria and archaea can grow in temperatures as hot as $\sim\SI{100}{\celsius}$ and $\sim\SI{122}{\celsius}$, respectively), this represents a significant loss of biodiversity and loss of most complex life.  This also roughly corresponds to the temperature regime past which precipitation in expected to decrease with increasing temperature \citep{liu24}.  As this mechanism is essential for the carbonate-silicate weathering feedback, this temperature would also correspond to the breakdown of the feedback cycle.  In addition, the behavior of precipitation cycles themselves may be significantly altered past this threshold, possibly resulting in periodic episodes of extremely high rainfall rates \citep{seeley2021episodic}.
This threshold corresponds to 0.11 bar $p$\ce{CO2} and a total ocean + atmospheric carbon inventory of $\sim\SI{458000}{GtC}$ ($11\times$ increase) for present-day solar luminosity.

\subsubsection{Hard Snowball States}
\label{sec:snowballs}
Hard snowball states occur when the entirety of a planet's oceans become frozen over at the surface. These are distinct from `waterbelt' or `slushball' states, where oceans near the equator may remain unfrozen. Snowball states can occur when GHG concentrations and subsequently global temperatures become low enough to trigger a runaway ice-albedo feedback \citep{Pierrehumbert2011}. While the cause of Earth's snowball events is contentious, it has been suggested that snowball states may be initiated by the rapid introduction of low-latitude mountain belts that increase global weathering rates (e.g., \citealt{Joshi2019}). 

Water ice reflects $\sim60\%$ of sunlight, more than land ($\sim30\%$) or open oceans ($\sim10\%$) \citep[e.g.,][]{haqq2016,Kadoya2019}. When climate cools and polar ice forms, reflection of sunlight by ice causes further surface cooling which in turn leads to more ice. This can eventually lead to total glaciation of the surface. Evidence suggests that Earth escaped such snowball states \citep{Kirschvink1992}; freezing soil curtails weathering, allowing for buildup of atmospheric \ce{CO2} via volcanism, eventually warming the planet enough to melt the ice. If the poles become cold enough for \ce{CO2} to condense, the snowball would likely become inescapable, although this effect is likely only important for planets that receive significantly ($\lesssim0.62 S_{\oplus}$) less instellation than modern-day Earth \citep{Pierrehumbert2011,TURBET201711}.

In our simulations, we treat the hard snowball case as a `climate catastrophe'. While the Earth has likely survived at least two clusters of snowball events (it is still unclear whether these were hard or waterbelt snowball states), including these events helps us better represent the \textit{overall} climate stability of varying tectonic setups.  In an astrobiological point-of-view, a hard snowball would almost certainly prevent the detection of molecular biosignatures due to life being trapped under the ice. Moreover, modeling exactly \textit{how} the climate escapes a Snowball Earth episode is still an active area of research (e.g., \citealt{godderis11,Kadoya2019}). 

The exact $p$\ce{CO2}/temperature threshold for initiating a runaway ice-albedo feedback is still debated and is sensitive to a suite of tunable (largely unconstrained) parameters, including snow/ice albedos, heat transport efficiency during a highly-glaciated state, and cloud feedbacks.  \citet{Pierrehumbert2011} show that in a simple 0-D model, temperatures as high as \SI{7}{\celsius} could initiate runaway glaciation, a threshold used by \citet{wordsworth21} and \citet{baum2022snowball}.  3-D general circulation model (GCM) simulations from \citet{Pierrehumbert2011} suggest $p$\ce{CO2} values of $\sim1000-\SI{2000}{\micro bar}$ (at $94\%\, L_{\odot}$) initiate a hard snowball, whereas \citet{yang2012} find that values as low as \SI{18}{\micro bar} are needed.
Studies of the cycling between hard snowball and deglaciated states find runaway initiation global-average temperatures of $\sim\SI{-8}{\celsius}$ \citep{Paradise2017,Kadoya2019} to $\sim\SI{1}{\celsius}$ \citep{menou2015}.

Due to the above discrepancies, we use a snowball initiation global temperature of \SI{0}{\celsius}, corresponding to the freezing point of surface water.  We note that, even with this seemingly high threshold temperature, we expect to underestimate the amount of snowball events due to the lack of ice-albedo feedback in our $\overline{T}(p\ce{CO2})$ relationship, where \SI{0}{\celsius} corresponds to \SI{11}{\micro bar} $p$\ce{CO2} and a total ocean + atmospheric carbon inventory of $\sim\SI{20000}{GtC}$ ($\sim2\times$ decrease).


\subsection{Trial Parameters}\label{sec:params}

We consider effects of overall land area coverage and number of disks on climate stability. We vary total land area as $A_{land,\bigoplus}/10$, $A_{land,\bigoplus}/3$, $A_{land,\bigoplus}/2$, $A_{land,\bigoplus}$, $3A_{land,\bigoplus}/2$, and $2A_{land,\bigoplus}$, where $A_{land,\bigoplus}=\SI{1.5e14}{m^2}$, approximating Earth's modern-day land coverage.  These land fractions correspond to 3\%, 10\%, 15\%, 30\%, 45\%, and 60\% surface area land fraction. We vary the number of disks as 2, 4, 8, and 12, defining a 24-member grid of scenarios. The $A_{land}=A_{land,\bigoplus}, N_{disk}=8$ scenario represents the `Earth-like' case. We run each trial for four billion years, comparable to the length of Earth's history of surface water \citep{Wilde2001,watson05} and possibly life \citep{Schopf2018}.

\section{Results} \label{ch:4}
\subsection{Catastrophically Low Survival} \label{sec:lowsurv}

\begin{figure}
    \centering
    \includegraphics[height=0.9\textheight]{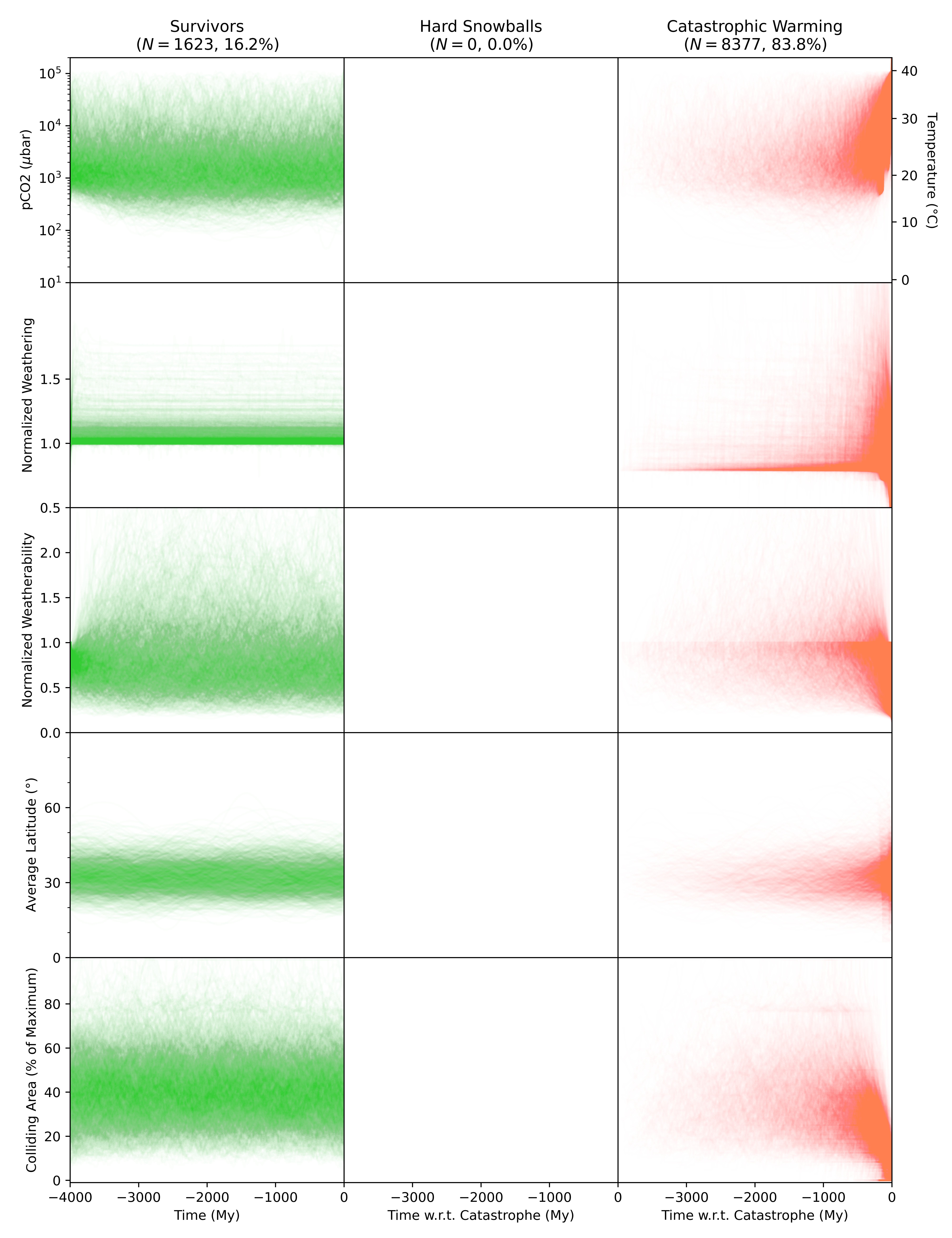}
    \caption{Output ($p$\ce{CO2}/average surface temperature, global weathering, weatherability, average land latitude, and colliding area) for 10000 model runs of the collisional case, where soil age is solely controlled by disk collisions (Section \ref{sec:lowsurv}).  In these model runs, frequent disk collisions are required to generate the high resurfacing rates needed to stabilize the climate over geologic timescales.  All trials that begin with no colliding disks quickly reach catastrophic warming as their soils become leached of cations available for weathering. Differences in the time-to-catastrophe for these trials is dictated by their prescribed \ce{CO2} outgassing rates, which are generally proportional to trial land area.}
    \label{fig:nomaxtag}
\end{figure}

Results for our initial trials can be seen in Figure \ref{fig:nomaxtag}.
In these trials, which show an average survival rate of only $16\%$ over 4 Gy, the fraction of fresh minerals available for weathering (and thus $\tau_{s}$) is only reset by disk collisions. Low survival rates occur because both total \ce{CO2} drawdown due to weathering and the feedback climate strength $\alpha$ decrease drastically as soils age without collisions.  Thus, worlds require constant resupply of cations via collisions to keep the balance between weathering and outgassing stable.  Trials that began with no colliding disks died within $\lesssim1$ Gy due to this effect. Those that did survive generally contained a substantial amount of colliding area throughout the 4 Gy run but had extreme variability in their $p$\ce{CO2} values over time, suggesting that these climates are not particularly stable.  We conclude the climate of worlds where the amount of silicate weathering is \textit{only} determined by the MAC model, including $p$\ce{CO2} and $T$ dependencies of weathering reactions, are inherently not stable and (in most cases) require additional negative feedbacks to remain habitable over geologic timescales.  

\subsection{Inclusion of a Maximum Effective Soil Age}

\label{sec:tmax}

On Earth, several processes contribute to cation resupply that are independent of processes modeled in \citet{Maher2014} and are not included in our idealized model. Up to $\sim10\%$ of modern-day silicate weathering involves windblown dust, increasing to up to $\sim50\%$ in tectonically inactive regions \citep{Hilley2008}. Dust input aids weathering in tectonically inactive regions by resupplying cations required for Reaction \ref{rxn:cscyclenet} without tectonic activity.

Physical erosion causes the weathering front to descend into the underlying bedrock, exhuming fresh minerals. Physical erosion rates vary across the globe and are linked to tectonic uplift and precipitation.  Rates vary from \SI{4}{\micro\metre.yr^{-1}} in tectonically inactive regions like Sri Lanka to \SI{3000}{\micro\metre.yr^{-1}} in the Himalayas \citep{GRANGER2007,Portenga2011} and are inversely proportional to the soil residence time (or soil age, $\tau_{s}$). Evidence suggests that the presence of glaciers can also speed up local erosion rates, and that the existence of glacial-interglacial cycles may be important in resupplying cations for weathering \citep{riebe2004}.

\begin{figure}
    \centering
    \includegraphics[height=0.9\textheight]{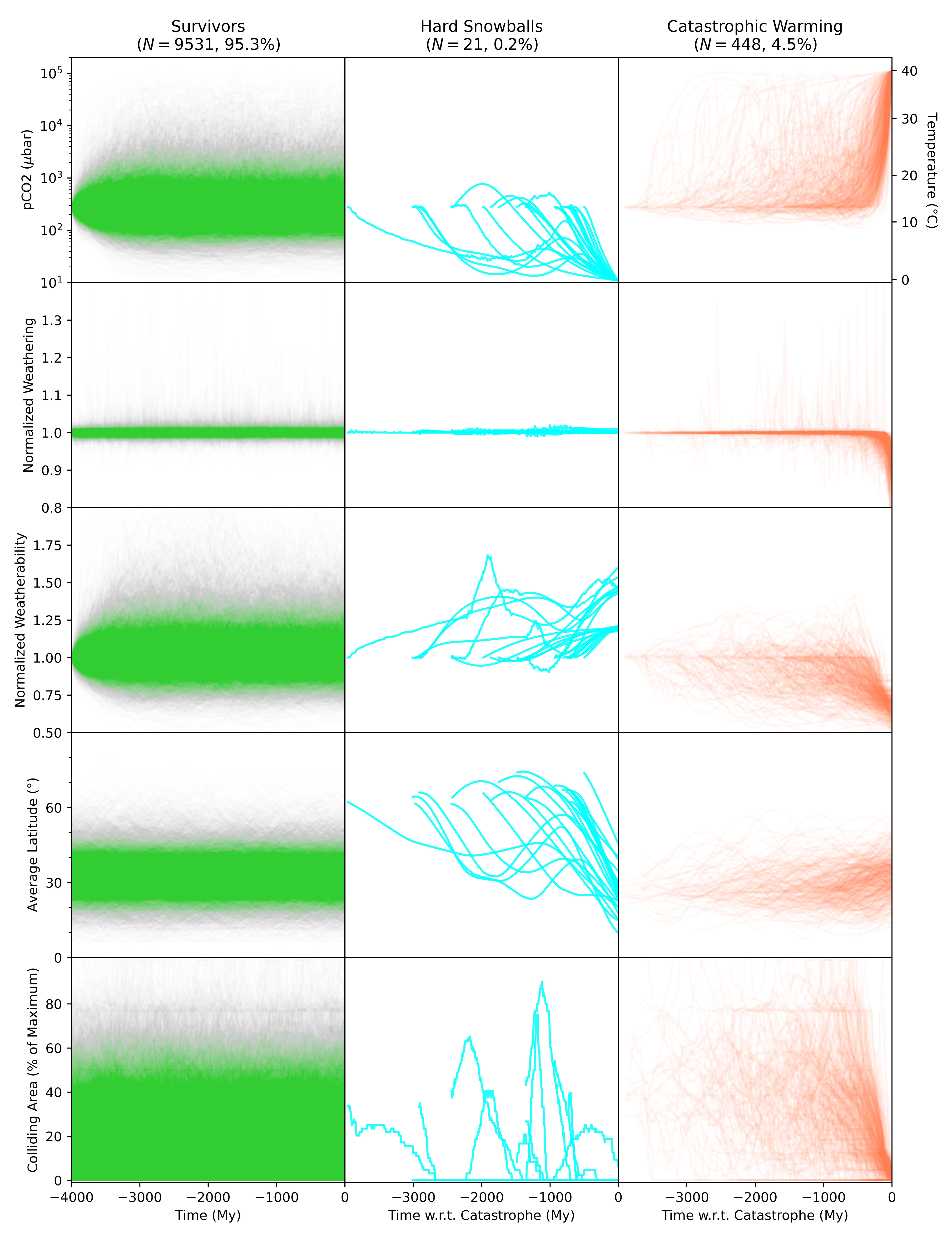}
    \caption{Output for 10000 trials where collisions reset the local effective soil age and soils retain a minimum fraction of fresh minerals for weathering approximating the effects of non-tectonic processes on weathering (Section \ref{sec:tmax}).  Despite large ($\gtrsim2\times$) perturbations in weatherability, the silicate weathering feedback is able to stabilize climate in the vast majority of trials. Most catastrophic warming events in these trials were triggered by previously-colliding disks breaking up, causing soils to age and weathering fluxes to decrease drastically. }
    \label{fig:wmaxt}
\end{figure}

We now apply a maximum effective soil age $\tau_{s,max}$, so that soils never become completely leached of weatherable minerals and always retain a minimum amount.  We assume that weathering in tectonically inactive regions corresponds to, on average, the `cratonic' value of $Dw=\SI{0.01}{m.yr^{-1}}$ defined in \citet{Maher2014}, based on river solute data of tectonic inactive regions. This yields $\tau_{s,max}\approx\SI{600}{kyr}$ and corresponding to a `fresh rock' fraction of $\sim2\%$. The inclusion of a maximum effective soil age increases the average survival rate over 4 Gy to $95\%$.  Aggregate results can be seen in Figure \ref{fig:wmaxt}.  Almost all of the catastrophic warming events resulted from trials with initially colliding disks that broke up over the course of the trial runtime---as continents that were previously colliding break-up, soils quickly become leached of cations, gradually reducing the weatherability, overall weathering \ce{CO2} drawdown, and the strength of the planet's overall weathering feedback.

Many of our catastrophic warming collision breakup events are accompanied by a continental drift polewards (shown by an increase in average latitude), which can also contribute to $p$\ce{CO2} buildup by drastically reducing the amount of precipitation available for weathering. These trials suggest that both frequent continental collisions \textit{and} an additional cation resupply mechanism are required for climate stability on geologic timescales.

\begin{figure}
\centering
\begin{subfigure}{.4\textwidth}
  \centering
  \includegraphics[width=\linewidth]{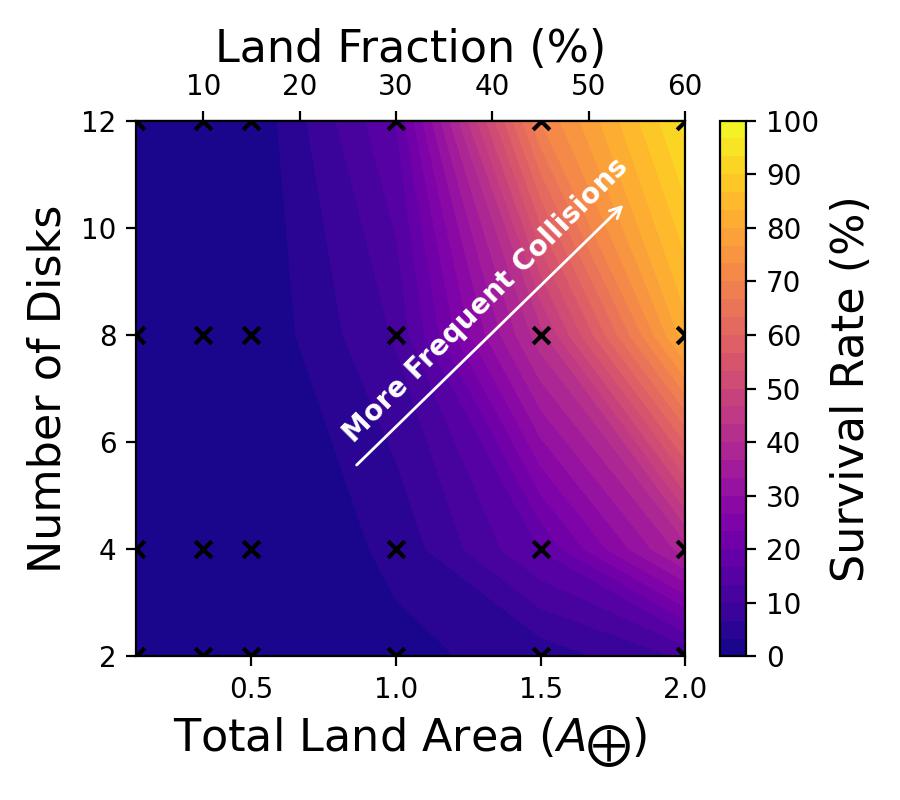}
  \caption{Old soils lose all weatherable minerals}
  \label{fig:sub1}
\end{subfigure}%
\begin{subfigure}{.4\textwidth}
  \centering
  \includegraphics[width=\linewidth]{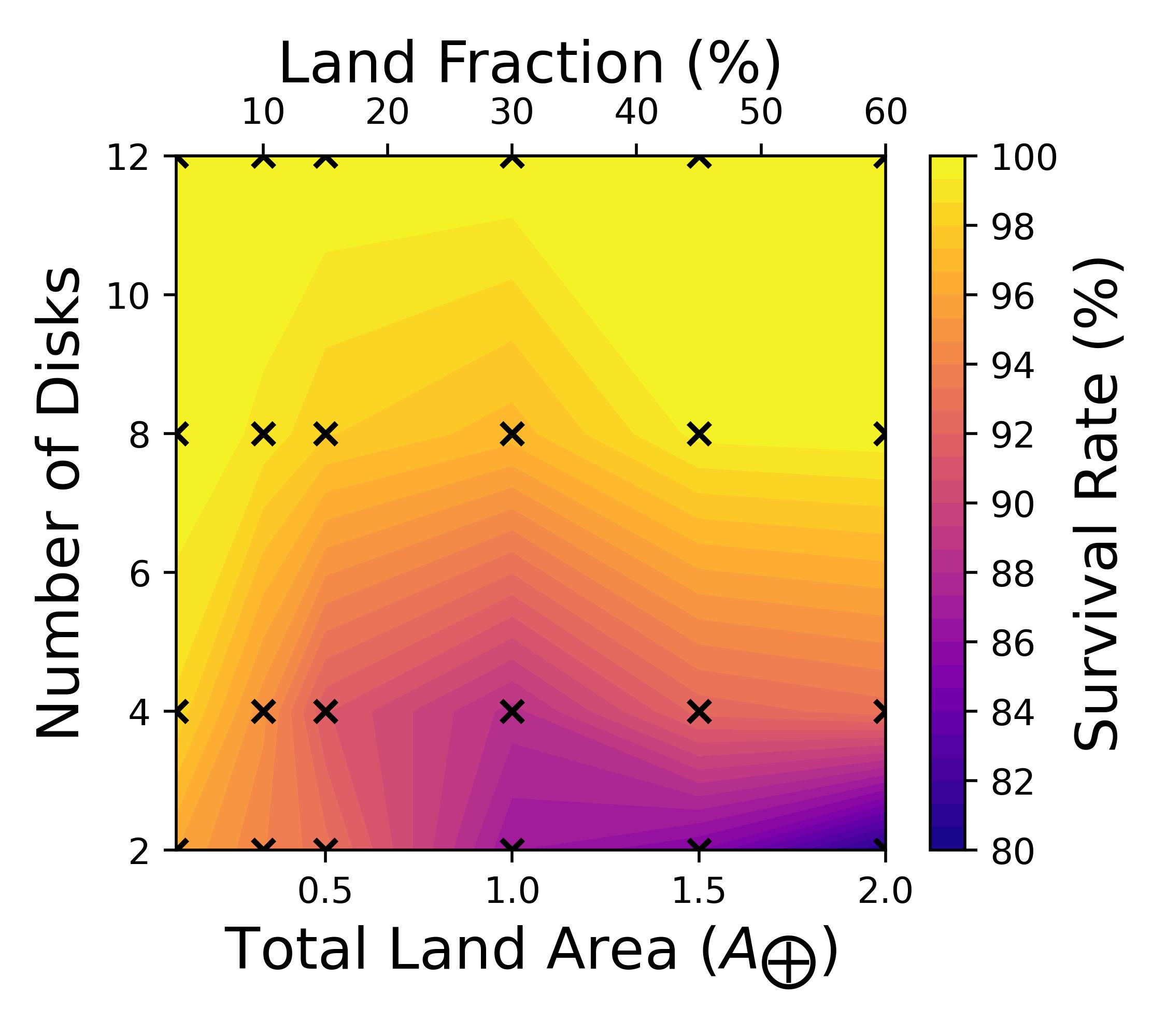}
  \caption{Old soils retain some fresh minerals}
  \label{fig:sub2}
\end{subfigure}
\caption{Survival rates for the 4 Gy collisional \texttt{DISKWORLD} model runs, where collisions form mountain belts endowed with young, highly weatherable soils.  Black crosses indicate the scenarios sampled.  
Survival rate for the Earth-like case ($A=A_{\bigoplus},N_{disk}=8$) is $10\%$ and $97\%$, respectively.}
\label{fig:survcontour}
\end{figure}

\subsection{Volcanic Outgassing Variation} \label{sec:fvolpert}
Volcanic outgassing ($F_{vol}$) has fluctuated throughout Earth's history [for compilations of Phanerozoic outgassing estimates, see \citet{Godderis2017} and \citet{Krissansen2017}].  The trials of Section \ref{sec:lowsurv}-Section \ref{sec:tmax} likely overstate the stability of runs with low amounts of outgassing where perturbations due to tectonic noise are few and far between due to the low amount of collisions in low $A_{land}$ trials.

We follow the approach of \citet{baum2022snowball} in modeling stochastic time-variability of $\ce{CO2}$ outgassing, where volcanic outgassing evolves as 

\begin{equation}
    \frac{dF_{vol}}{dt}=\frac{1}{\tau}(\mu_{vol}-F_{vol})+\sigma B(t),
\end{equation}
where $\mu_{vol}$ is a prescribed `average' outgassing flux, $\tau$ is the relaxation timescale, $\sigma$ represents the strength of variations, and $B(t)$ is a Gaussian noise term (with a mean of 0, standard deviation of 1) evaluated at each timestep. High values of $\tau$ represent \textit{longer-term} variations whereas high values of $\sigma$ represent \textit{stronger} variations.  For finite timesteps this is accomplished via an Euler–Maruyama method,
\begin{equation}
    \Delta V=\frac{1}{\tau}(\mu_{vol}-F_{vol})\Delta t+\sigma\mathcal{N}(0,1)\sqrt{\Delta t}.
\end{equation}
This approach assumes that outgassing tends towards a long-term mean $\mu_{vol}$ due to a negative feedback mechanism, which may not be the case.  Nevertheless, this approach is a reasonable starting point.  We scale the noise term by the starting outgassing flux, causing trials to experience the same \textit{fractional} variation, so that low $F_{vol}$ trials are not inherently more unstable,

\begin{equation}
   \sigma^{*}=\sigma\times\frac{\mu_{vol}}{\SI{0.084}{GtC.yr^{-1}}},
\end{equation}
where \SI{0.084}{GtC.yr^{-1}} (\SI{7}{Tmol.yr^{-1}}) is the value of $\mu_{vol}$ prescribed in \citet{baum2022snowball}. In our model, $\mu_{vol}$ is set by assuming $\mu_{vol}=F_{vol}(t=0)$.  

\begin{figure}[htp]

\begin{subfigure}{\textwidth}
\centering
\includegraphics[width=0.7\linewidth]{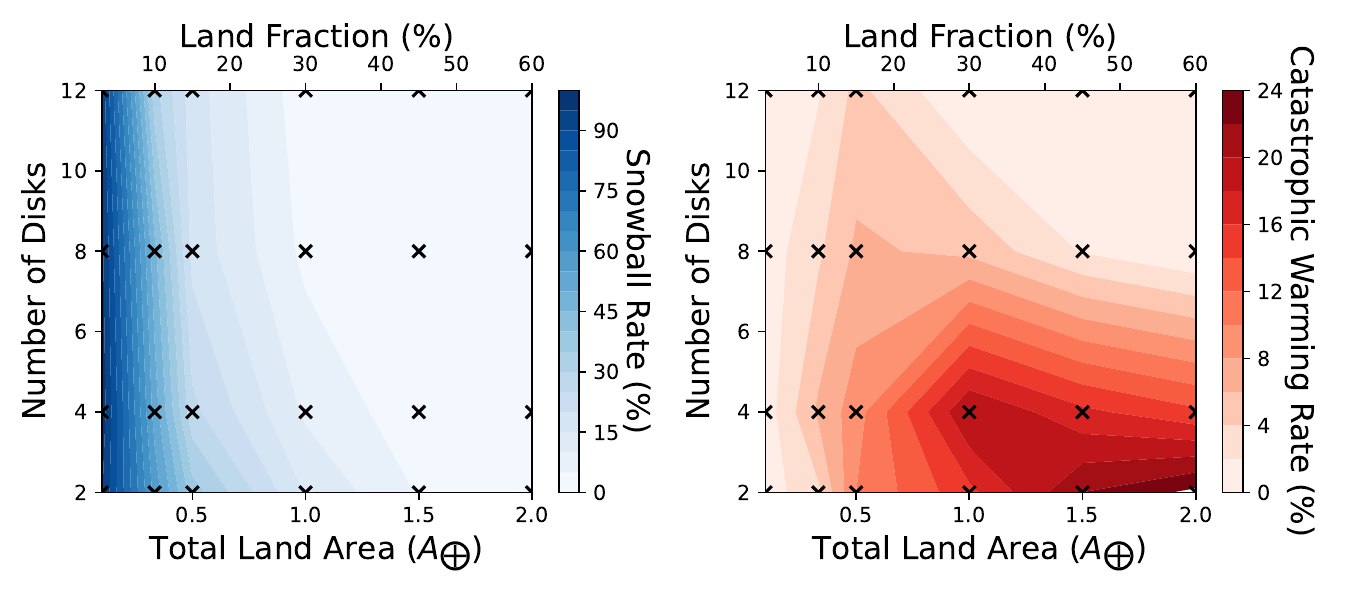}
\caption{Variable $A$ and $N$, $\tau=\SI{30}{Myr}, \sigma=\SI{2e-4}{Tmol.yr^{-1}}$}
\end{subfigure}


\begin{subfigure}{\textwidth}
\centering
\includegraphics[width=0.7\linewidth]{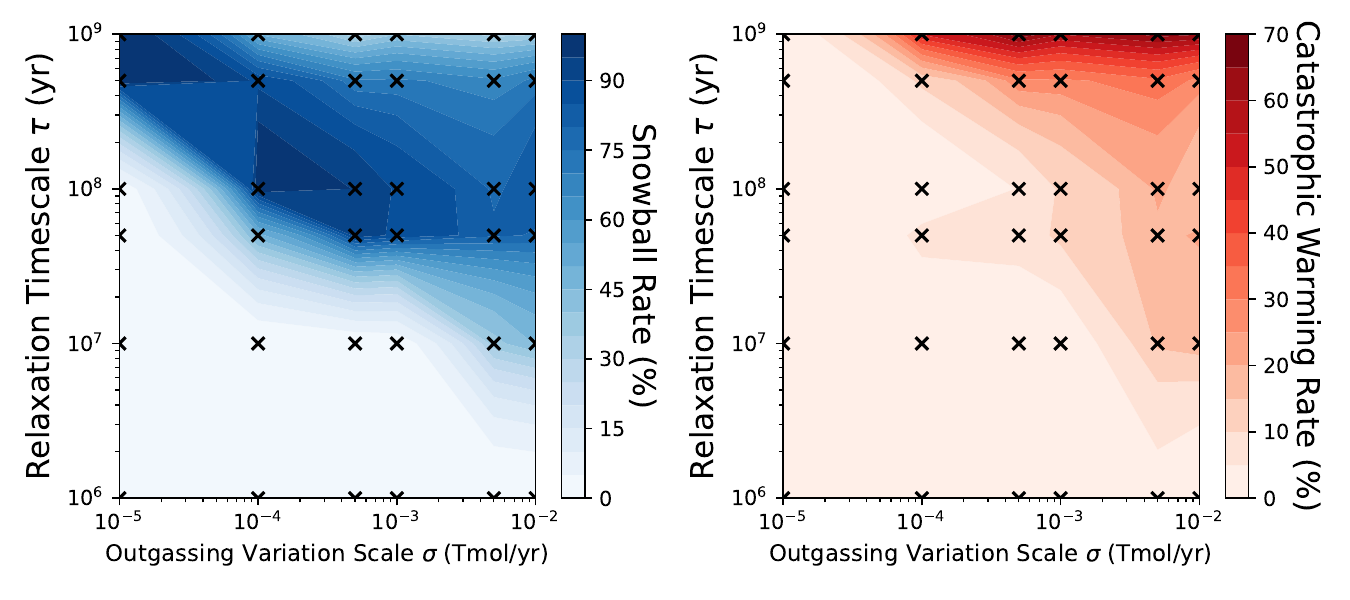}
\caption{Variable $\tau$ and $\sigma$, $A=A_{\oplus}, N=8$}
\end{subfigure}

\caption{Snowball and catastrophic warming rates over 4 Gyr for variable outgassing trials explained in Section \ref{sec:fvolpert}.  Crosses indicate scenarios sampled.  These trials reveal that planets with lower land fractions, while stable to tectonic noise, are highly susceptible to climate catastrophes caused by volcanic outgassing variability. In addition, longer-term and higher amplitude outgassing perturbations spell doom for an Earth-like  [$N=8, A_{land}=A_{land,\oplus}$, panel (b)] setup. Note that the apparent decrease in snowball rate at long relaxation timescales is caused by an accompanying increase in catastrophic warming, and thus the total survival rate does not increase.}\label{fig:voutgas}

\end{figure}

We test the effects of variable volcanic outgassing by both: (1) varying the number of disks and total land area as in previous trials and (2) varying the strength $\sigma$ and timescale $\tau$ of variations.  In scenario (1), we use `intermediate' values of $\tau=\SI{30}{Myr}$ and $\sigma=\SI{2.4e-6}{GtC.yr^{-1}}$ (\SI{2e-4}{Tmol.yr^{-1}}) from \citet{baum2022snowball}.   
Results can be seen in Figure \ref{fig:voutgas}.  These runs show that planets with low amounts of land area are particularly susceptible to variations in volcanic outgassing, due to the lack of tectonic collisions replenishing weatherable minerals and thus lack of a strong stabilizing feedback.  Furthermore, the strength and duration of variability have a profound effect on climate stability, with few trials surviving with $\tau\gtrsim\SI{50}{Myr}$ or $\sigma\gtrsim\SI{5e-3}{Tmol.yr^{-1}}$.  \citet{baum2022snowball} note that determining $\tau$ and $\sigma$ values appropriate for Earth is difficult and that these values also likely evolve with time, but that estimates of Phanerozoic outgassing rates hint that $\tau\gtrsim\SI{100}{Myr}$ for Earth.

Both scenarios show higher rates of snowball events than previous trials relative to catastrophic warming events. This is due to carbon depletion required to initiate a snowball ($\sim\SI{21000}{GtC}$) being much less than the carbon addition required to initiate catastrophic warming ($\sim\SI{420000}{GtC}$) with our assumption of $\overline{T}_{0}=\SI{13}{\celsius}$, thus causing more susceptibility to source-based perturbations.

\subsection{Faint Young Sun and Solar Luminosity} \label{sec:FYS}

Solar luminosity has risen by $\sim40\%$  over the last 4 Gy \citep{Bahcall_2001}. It is believed that GHG concentrations on early Earth must have been much higher to compensate for this `Faint Young Sun' paradox \citep{HART1978,walker1985,Charnay_2020}.  Despite a $\sim\SI{20}{\celsius}$ colder equilibrium temperature, it is thought that water has been present on Earth's surface since $\sim4.4$ Gya \citep{Wilde2001}, and that Earth has been ice-free for most of its history \citep{EVANS2003}. At low solar luminosity,  GHG concentrations much higher than present atmospheric levels would have been required to sustain ice-free conditions \citep{Charnay_2020}.  Whether this gas was \ce{CO2}, methane (\ce{CH4}), or a more exotic species is still up for debate.   

We investigate how \texttt{DISKWORLD}'s \ce{CO2} concentrations evolve with changing solar luminosity. In our model, increasing solar luminosity warming the surface would increase global weathering.  The increased weathering would drawdown \ce{CO2}, acting as a negative feedback to the initial surface warming. Thus, we would expect the atmospheric \ce{CO2} concentration to decrease over time.

Following \citet{Krissansen2017} (originally introduced in \citealt{gough1981}), we assume that solar luminosity evolves with time as:
\begin{equation}\label{eq:luminosity}
    L(t)=\frac{L_{\bigodot}}{1+\frac{2}{5}(1-\frac{t}{4.57\textrm{ Gy}})},
\end{equation}
where $L_{\bigodot}$ is present-day solar luminosity. We test the effects of changing luminosity on $p$\ce{CO2} and temperature over time. We prescribe an initial start time of 3 Gya ($t=\SI{1.57}{Gy})$, corresponding to a starting luminosity of $0.79L_{\odot}$.

We solve for initial conditions assuming $T(t=0)=\SI{13}{\celsius}$ as with previous trials, leading to an initial $p$\ce{CO2} of \SI{86.5}{mbar} and an ocean reservoir of \SI{321050}{GtC}. The evolution of geothermal heat flow and \ce{CO2} outgassing fluxes is modeled after \citet{Krissansen2018}. In their framework, the \textit{relative} geothermal heat flow $Q$ changes over time as
\begin{equation}
    Q(t)=t^{-n_{out}},
\end{equation}
where $n_{out}$ is an assumed exponential dependence.  \ce{CO2} outgassing is related to $Q$ by 
\begin{equation}
\frac{F_{vol}}{F_{vol,0}}=\left(\frac{Q}{Q_{0}}\right)^{m},
\end{equation}
where $F_{vol,0}$ is the initial outgassing flux, $Q_{0}$ is the initial geothermal heat flux, and $m$ is another assumed exponential dependence.  Here, we adopt the median values of $n_{out}$ and $m$ used in \citet{Krissansen2018} (0.365 and 1.5, respectively), noting that these values are poorly constrained for the Earth.  These values cause \ce{CO2} outgassing to decrease by $\sim 44\%$ over 3 Gy.

\begin{figure}
    \centering
    \includegraphics[width=0.9\linewidth]{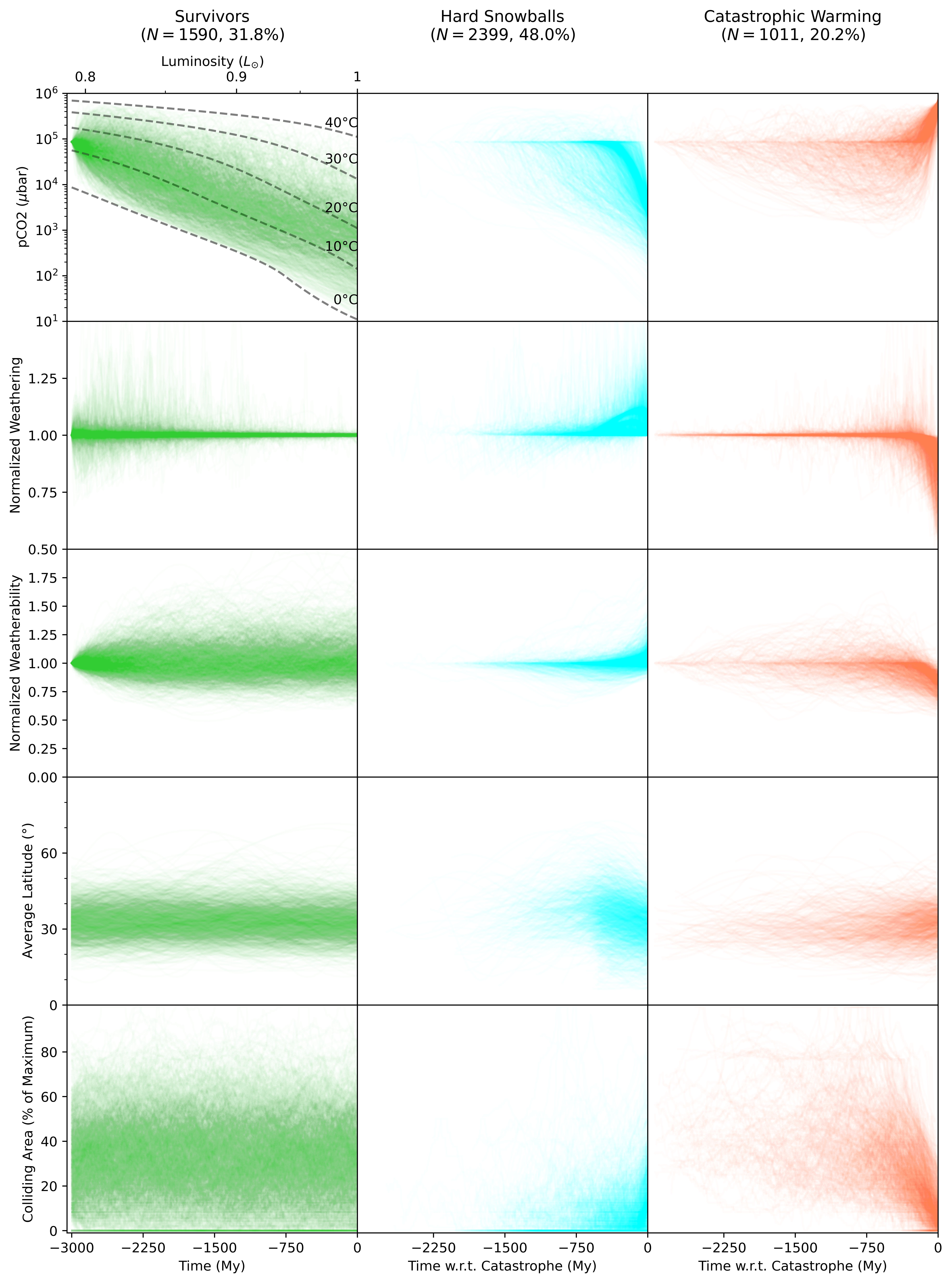}
    \caption{Output for 5000 trials including evolving solar luminosity and outgassing fluxes as described in Section \ref{sec:FYS}.  }
    \label{fig:FYS}
\end{figure}

\begin{figure}
    \centering
    \includegraphics[width=0.5\linewidth]{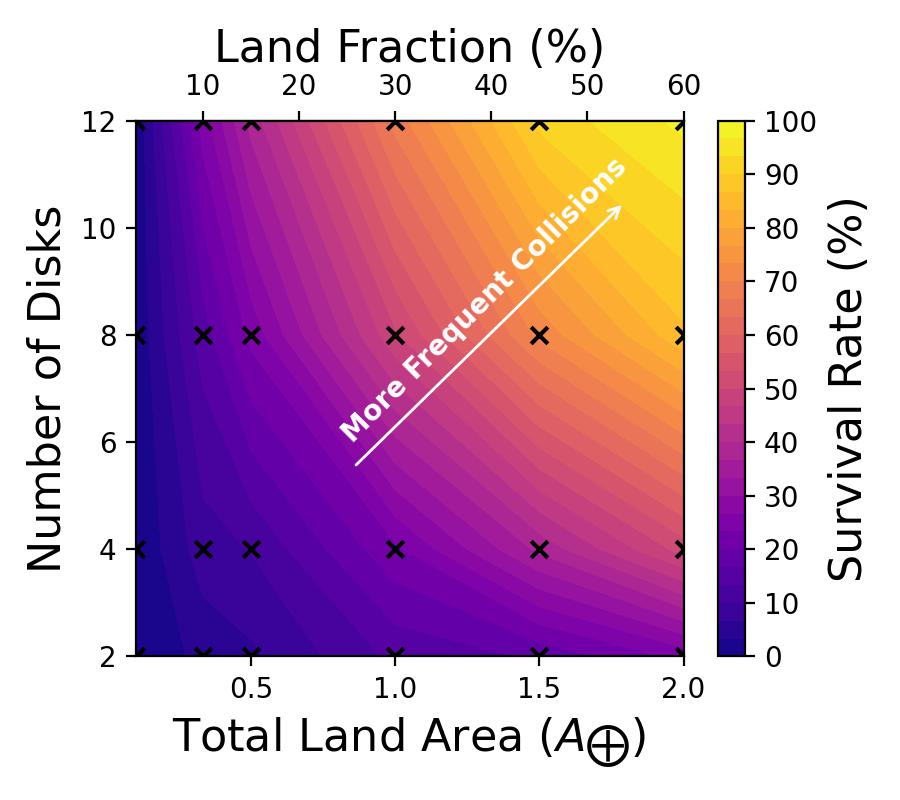}
    \caption{Survival rates over 4 Gyr as a function of disk number and total land area for trials with evolving solar luminosity and evolving outgassing [based on geothermal heat flux] (Section \ref{sec:FYS}). Low survival rates for both low number of disks and low land area augment similar results from varying outgassing fluxes Figure \ref{fig:voutgas}, and suggest that frequent collisions are required for long-term habitability.}
    \label{fig:FYS_surv}
\end{figure}

Model results for evolving solar luminosity can be seen in Figure \ref{fig:FYS}.  These trials show a stark contrast with our fixed luminosity trials of Section \ref{sec:tmax}; survival rate is only 32\% despite including a maximum effective soil age.  Trials that do survive require continuous frequent collisions in order to replenish minerals available for weathering, reinforcing our previous results. The high rate of snowball catastrophes (48\%) is associated with trials that begin with a low amount of colliding land and experience short-term collisions augmented by latitudinal drift towards the equator, both effects increasing weathering rates substantially.  Catastrophic warming events correspond to short-term break-up of a large area of colliding disks, causing weathering rates to plummet and \ce{CO2} to gradually rise.  Survival rates as a function of disk number and land fraction are shown in Figure \ref{fig:FYS_surv}.

\begin{figure}
    \centering   
    \includegraphics[width=0.9\linewidth]{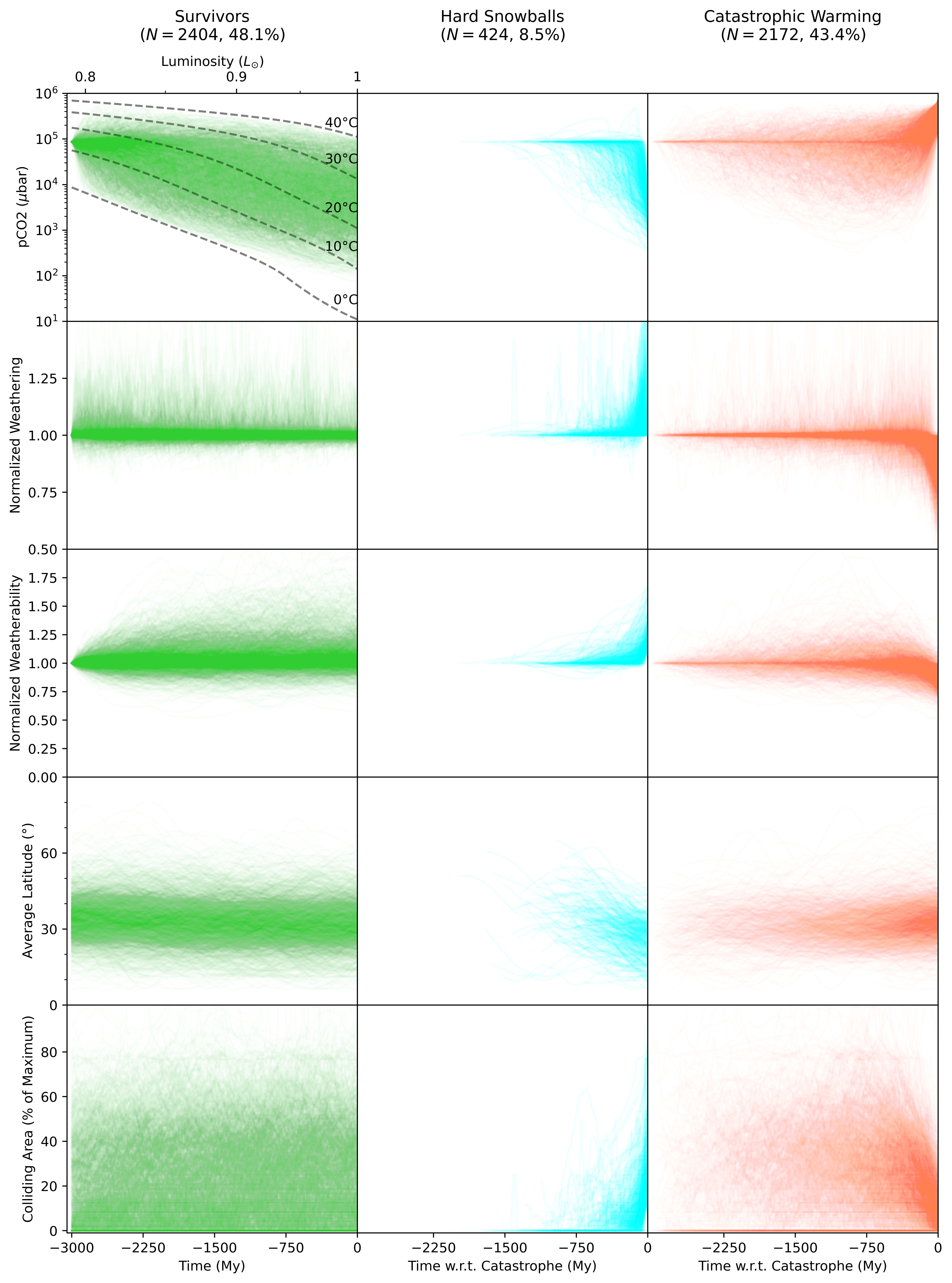}
    \caption{Output for 5000 trials including evolving solar luminosity with a fixed outgassing flux. High \ce{pCO2} initial states at lower luminosities lead to high initial weathering and outgassing rates.  If these outgassing rates do not decrease due to evolving geothermal heat fluxes, the carbonate-silicate weathering feedback is not strong enough to prevent catastrophic warming in a large number of these trials.}
    \label{fig:FYS_noevol}
\end{figure}

As the long-term evolution of \ce{CO2} is poorly-constrained, we also explore the case where \ce{CO2} outgassing remains constant over time while luminosity evolves, with results shown in Figure \ref{fig:FYS_noevol}.  Similar to Figure \ref{fig:FYS_surv}, many more catastrophic warming events occur and surviving trials clustering around much higher global temperatures.  This is due to the $p$\ce{CO2}-dependencies of weathering (Equation \ref{eq:pco2dep}) causing weathering fluxes to gradually decrease with increasing luminosity for a given temperature, causing long-term imbalances in weathering and outgassing \ce{CO2} fluxes. Surviving trials again require frequent continental collisions to remain stable over the model runtime, and many large-scale breakup events trigger catastrophic warming, supporting our results including geothermal heat flux evolution.

The high rate of climate catastrophes may be due to simplifying assumptions made, including ocean chemistry, how rainfall rates behave in high $p$\ce{CO2} \& low luminosity climates, the evolution of outgassing rates, and bulk atmospheric composition.
It is difficult to qualitatively assess the effects of increasing luminosity without a coupled model of the time-evolution of (1) \ce{CO2} outgassing, (2) land fraction, and (3) ocean chemistry, which are not well-constrained even for Earth \citep{korenaga13,Krissansen2017,Krissansen2018,krissansen20}. However, our model results support the notion that worlds with few continental plates or low land fraction are particularly susceptible to climate catastrophes.

\subsection{Temperature/$p$\ce{CO2}-Dependent Seafloor Weathering}\label{sec:sfw}

Recent models have strongly suggested that seafloor weathering (SFW) exhibits a temperature and pH dependence that acts as an additional stabilizing feedback to the slow carbon cycle \citep{Krissansen2017,Krissansen2018}. In this model, the rate of seafloor weathering $W_{sf}$ depends on the pH ($\ce{pH}=-\log_{10}{[\ce{H+}]}$) and temperature of the pore-space where dissolution occurs,
\begin{equation} \label{eq:KTsfw}
    W_{\ce{CO2},sf}\propto r_{spread-rate}\, e^{\frac{-E_{bas}}{RT_{pore}}} \, [\ce{H+}]_{pore}^{\gamma}.
\end{equation}

Here $r_{spread-rate}$ is the rate of seafloor spreading introducing new minerals available for weathering, $\gamma$ represents the pH dependence of seafloor weathering, $E_{bas}$ is the apparent activation energy for basalt dissolution (note that this is different from the quantity $E_{a}$), and $T_{pore}$ is the deep water temperature in the pores where dissolution is occurring.   We can directly calculate the pH of the ocean at any given time following methods in Appendix \ref{sec:csys}. Following \citet{Krissansen2018}, we assume $T_{pore}$ (in K) is linearly related to the average global surface temperature,
\begin{equation} 
    T_{pore}=1.02\times \overline{T}-\SI{16.7}{K}.
\end{equation}
Seafloor weathering in our model is assumed to cease when the pore temperature is below the freezing point of seawater, assumed to be \SI{-2}{\celsius} (corresponding to $\overline{T}=\SI{9}{\celsius}$).  For stability purposes, we ensure $W_{\ce{CO2},sf}$ is continuous by adding an additional $(T_{pore}-\SI{271}{\kelvin})/(\SI{2}{\kelvin})$ multiplicative term if $T_{pore}$ is below \SI{0}{\celsius}.

Using median values of $\gamma=.25$ and $E_{bas}=\SI{80}{kJ\per mol}$ from \citet{Krissansen2018} \citep[following][]{hayworth2020}, we obtain a climate feedback strength of $\alpha=\SI{16.2}{\%.\celsius^{-1}}$ at $T=\SI{12}{\celsius}$. Simpler models of seafloor weathering used to constrain $p$\ce{CO2} values for ancient Earth from \citet{sleep2001}, where
\begin{equation}\label{eq:sleepsfw}
W_{\ce{CO2},sf}\propto p\ce{CO2}^{0.7},   
\end{equation}
show a similar climate feedback strength of \SI{15.2}{\%.\celsius^{-1}}.

These high values of $\alpha$ imply that seafloor weathering fluxes are much more sensitive, and thus more climate-stabilizing, to perturbations in global temperature than continental silicate weathering. While continental weathering likely dominates the silicate weathering feedback on modern-day Earth, the stabilizing effect of seafloor weathering would be stronger for worlds with a higher global ocean fraction, assuming active tectonic resurfacing on the ocean floor \citep{lee2018,Lee2019}. This would help stabilize worlds that see few continental plate collisions. While many models infer relationships based on atmospheric $p$\ce{CO2} and proxies for ocean pH, which are also heavily affected by other unrelated processes, isotopic analyses have begun to constrain temperature and $p$\ce{CO2} dependencies of SFW (e.g., \citealt{coogan2015,coogan2018,coogan2022}). However, it remains unclear how Equation \ref{eq:KTsfw} behaves at very high or low values of $p$\ce{CO2} relevant to our end-of-habitability cases or during the distant past where solar luminosity was significantly lower.

\begin{figure}
    \centering
    \includegraphics[width=0.8\linewidth]{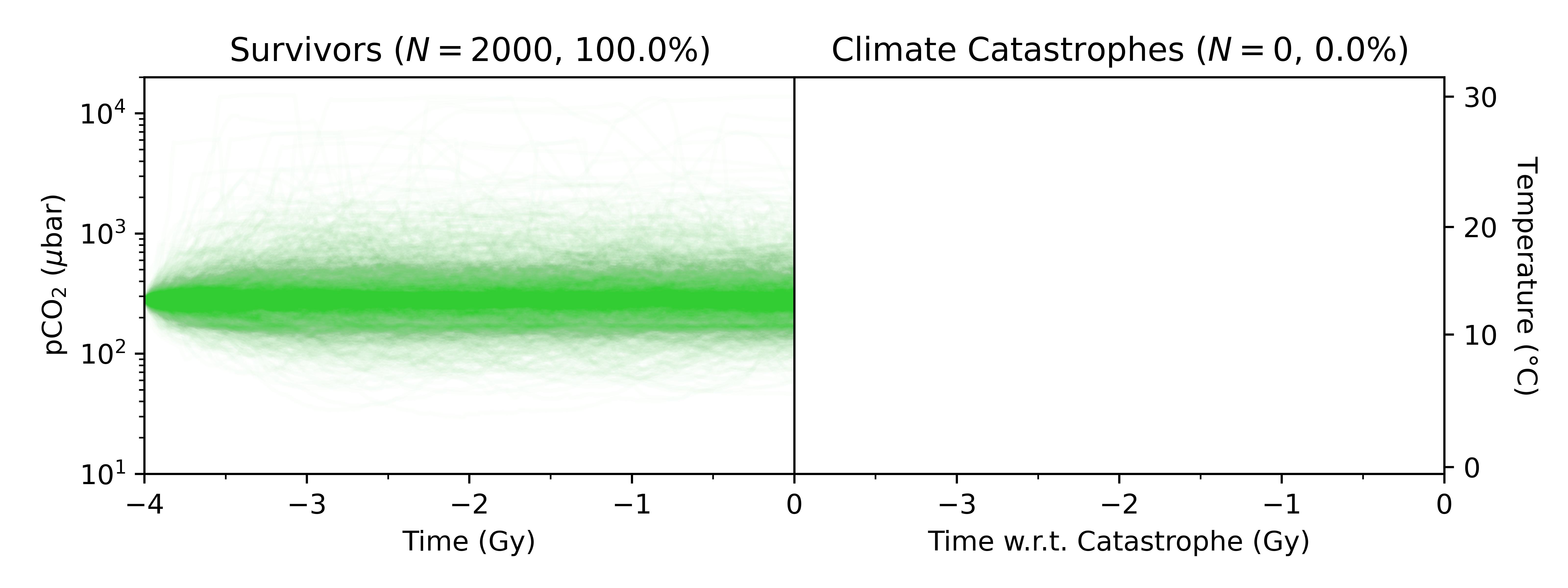}
    \caption{Model runs showing the time variability of $p$\ce{CO2} for a wide variety of tectonic setups, including $p$\ce{CO2}/temperature-dependent seafloor weathering as discussed in Section \ref{sec:sfw} based on \citet{Krissansen2018}, omitting luminosity evolution and outgassing variation.  In these trials, seafloor weathering acts as an extremely stabilizing negative feedback, showing only small variations in $p$\ce{CO2} over Gy timescales, and suggest that seafloor weathering may be an additional important stabilizing negative feedback in the slow carbon cycle.}
    \label{fig:sfw_pco2}
\end{figure}

We test the effects of including $p$\ce{CO2}/$T$-dependent seafloor weathering in \texttt{DISKWORLD}. In these trials, we replace our constant SFW flux with SFW modeled after \citet{Krissansen2018} and \citet{hayworth2020}.  We assume that the rate of seafloor spreading remains constant over time, i.e.
\begin{equation}
    W_{\ce{CO2},sf}=\SI{0.0054}{GtC.yr^{-1}}\times e^{\frac{-E_{bas}}{RT_{pore}}}\, [\ce{H+}]_{pore}^{\gamma}.
\end{equation}
We find that the relationship between weathering and global temperatures/$p$\ce{CO2} described in Equation \ref{eq:KTsfw} is \textit{extremely} efficient in correcting imbalances between weathering and outgassing fluxes.  Over 2000 trials varying land area and distribution, we record zero climate catastrophes.  In addition to having no climate catastrophes, $p$\ce{CO2} values remained stable for most trials. As shown in Figure \ref{fig:sfw_pco2}, the maximum $p$\ce{CO2} reached across these trials is \SI{14.3}{mbar} ($\overline{T}=\SI{30}{\celsius}$), far from our end-of-habitability threshold.

\begin{figure}
    \centering
    \includegraphics[width=0.9\linewidth]{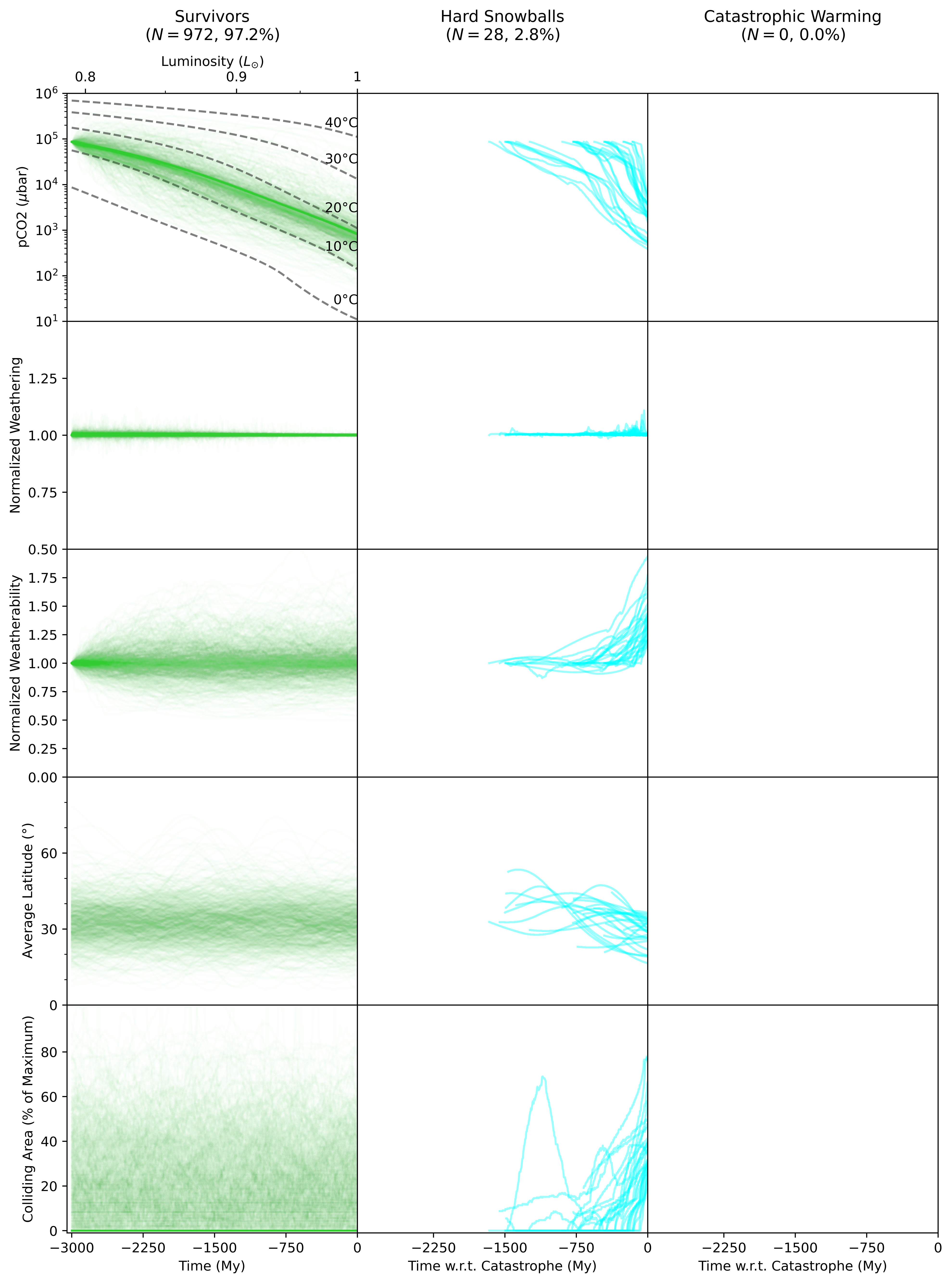}
    \caption{Model output for 1000 trials with evolving solar luminosity including $p$\ce{CO2}/temperature-dependant seafloor weathering as described in Section \ref{sec:sfw}. Similar to Figure \ref{fig:sfw_pco2}, these trials, when compared to the low survival rate of Figure \ref{fig:FYS}, further suggest that seafloor weathering acts as an additional important stabilizing feedback to the slow carbon system.  Climate catastrophes in these trials are exclusively caused by the sudden introduction of large amounts of weatherable material at low latitudes.}
    \label{fig:FYS_sfw}
\end{figure}

Including the effects of evolving solar luminosity as outlined in Section \ref{sec:FYS} further support $p$\ce{CO2}/temperature-dependant SFW as a highly stabilizing feedback.  In these trials, in addition to evolving solar luminosity and outgassing fluxes, we assume that the seafloor-spreading rate $r_{spread-rate}$ (Equation \ref{eq:KTsfw}) evolves proportional to geothermal heat flux $Q$, so that the base SFW rate is $48\%$ higher at the start of these trials (3 Gya).  Results can be seen in Figure \ref{fig:FYS_sfw}. Similar to Figure \ref{fig:sfw_pco2}, $p$\ce{CO2}-dependant SFW acts as an extremely stabilizing negative feedback, with few trials experiencing \textit{any} significant warming or cooling, and increasing the aggregate survival rate from 32\% to 97\%.

These results emphasize the need for further study exploring the direct $p$\ce{CO2} and temperature dependencies of seafloor weathering. The high climate feedback strength values predicted by previous studies (\citealt{sleep2001,Krissansen2018}, Equations \ref{eq:sleepsfw} and \ref{eq:KTsfw}) when compared to that of continental weathering suggest that seafloor weathering may even be the primary regulator of Earth's modern climate.   As with WHAK, the simplified approach of Equation \ref{eq:KTsfw} does not account for the depletion of cations available for weathering on the seafloor over time, although it does account for the introduction of fresh rock via seafloor spreading.


\subsection{Influence of Starting Temperature and $p$\ce{CO2} on Model Outcome}

\begin{figure}
    \centering
    \includegraphics[width=0.6\linewidth]{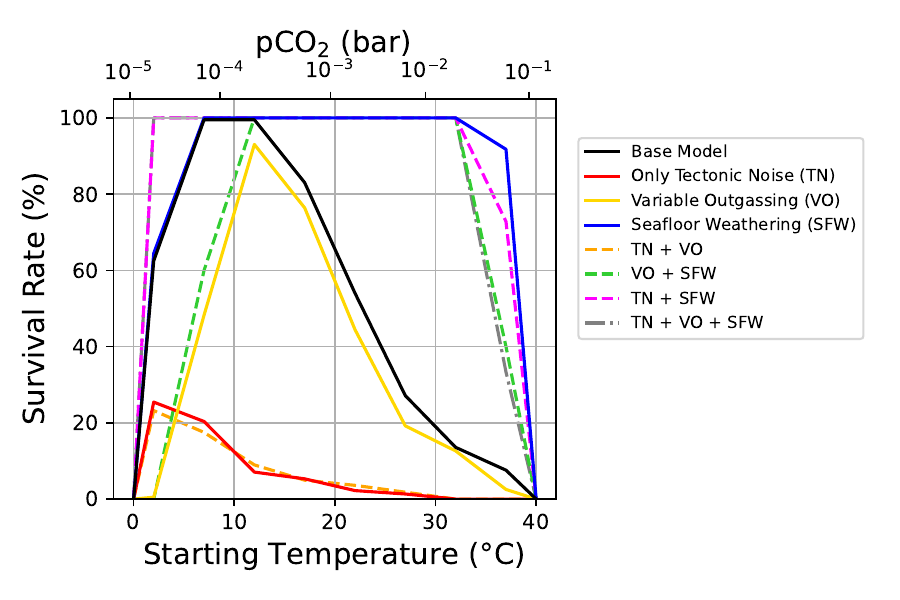}
    \caption{Survival rates over 4 Gyr for an Earth-like tectonic setup ($A_{land}=A_{land,\oplus}, N=8$) with a variety of starting global average temperatures assuming present-day solar luminosity.  `Seafloor Weathering' trials include $p$\ce{CO2}/temperature-dependent seafloor weathering as discussed in Section \ref{sec:sfw}, `Only Tectonic Noise' trials do not include the maximum soil age discussed in Section \ref{sec:tmax}, and `Variable Outgassing' trials use the `moderate' values for stochastic outgassing variation ($\tau=\SI{30}{Myr}$, $\sigma=\SI{2e-4}{Tmol.yr^{-1}}$, Section \ref{sec:fvolpert}).}
    \label{fig:tcurve}
\end{figure}

So far, we assumed a starting temperature similar to that of modern-day Earth.  Now we relax this assumption, computing survival rates for our Earth-like ($N=8, A_{land}=A_{land,\oplus}$) setup, including processes discussed in Section \ref{sec:lowsurv}, \ref{sec:fvolpert}, \& \ref{sec:sfw} not included in our base model.  We consider starting temperatures of [2, 7, 12, 17, 22, 27, 32, 37]\si{\celsius}, with results shown in Figure \ref{fig:tcurve}.  Stochastically varying outgassing flux has a major effect on survival rate at low starting temperatures and thus low $p$\ce{CO2}.  Our results augment similar studies concluding that low-$p$\ce{CO2} climates, like those expected near the inner edge of the habitable zone, are inherently less stable to given external perturbations to the carbonate-silicate weathering feedback (e.g., \citealt{wordsworth21,graham21}). Trials that only allow for soil age to be affected by tectonic collisions (`Only Tectonic Noise', Section \ref{sec:lowsurv}) remain highly susceptible to climate catastrophes, with very few trials surviving over 4 Gyr.

$p$\ce{CO2}/temperature-dependant seafloor weathering robustly stabilizes climate at all starting temperatures when including a variety of destabilizing processes, further emphasizing the need for more research on this process.  Almost no trials including seafloor weathering with starting temperatures of $\sim7-\SI{32}{\celsius}$ experience a climate catastrophe.


\section{Discussion} \label{sec:disc}

\subsection{Lack of Hard Snowballs}
Very few trials without outgassing variations (Section \ref{sec:lowsurv} and \ref{sec:tmax}) with fixed present-day solar luminosity ended in the snowball climate catastrophe described in Section \ref{sec:snowballs}.  While evidence suggests that multiple snowball events have occurred on Earth in the past (e.g., \citealt{Hoffman2017}), our model's behavior is not unexpected. As solar luminosity increases, the amount of \ce{CO2} required to trigger a runaway glaciation of the Earth's surface decreases significantly. The value of $p$\ce{CO2} at which this runaway occurs greatly affects the likelihood of our trials resulting in a snowball climate catastrophe. Even in highly-aged soils, the weathering feedback strength $\alpha$ is strong enough to prevent $p$\ce{CO2} reaching the snowball threshold at present-day luminosity ($p$\ce{CO2}=\SI{11}{\micro bar}). Climate feedback strength ($\alpha$) in the MAC model increases drastically as the global average surface temperature $\overline{T}$ decreases (Figure \ref{fig:feedback}). For mountainous soils at the equator, $\alpha$ increases by $39\%$ for a temperature change $\SI{15}{\celsius}\rightarrow\SI{5}{\celsius}$.  This value increases to $192\%$ for cratonic soils. This means that, in our model, colder climates are more efficient in rebalancing weathering and outgassing fluxes.

\begin{figure}
    \centering
    \includegraphics[width=0.7\linewidth]{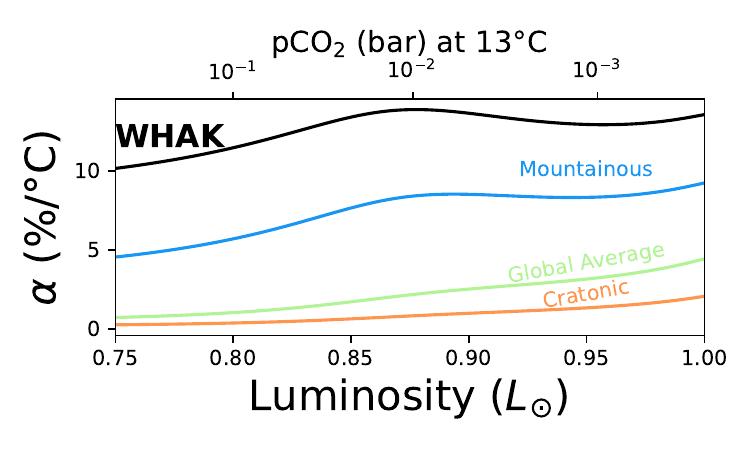}
    \caption{Climate feedback strength $\alpha$ as a function of solar luminosity for a fixed global average temperature of $\overline{T}=\SI{13}{\celsius}$ (assuming $0\degree$ latitude). For a fixed global temperature, lower luminosity entails higher $p$\ce{CO2} concentrations. Climates with higher $p$\ce{CO2} show a lower climate feedback strength due to the $p$\ce{CO2} dependency of weathering in the WHAK and MAC models. 
    }
    \label{fig:feedback_lum}
\end{figure}

\begin{figure}
    \centering
    \includegraphics[width=0.9\linewidth]{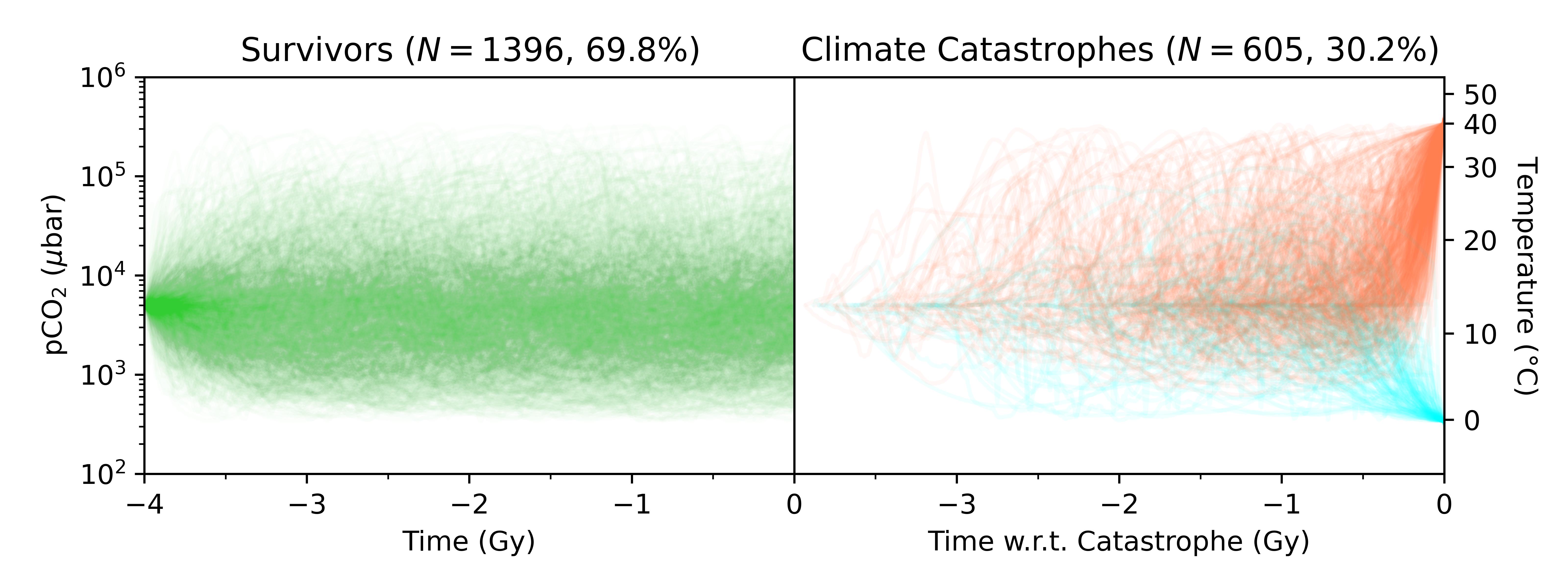}
    \caption{Model results from 2000 trials assuming a fixed luminosity of $L=0.9L_{\odot}$. These trials show much higher catastrophe rates than trials run at modern-day luminosity (Section \ref{sec:tmax}, snowballs: 6\% vs. 0.2\%, catastrophic warming: 23\% vs. 4.5\%) due to weaker climate feedback strength in high $p$\ce{CO2}-climates (Figure \ref{fig:feedback_lum}).}
    \label{fig:lowL}
\end{figure}

In addition, the climate feedback strength $\alpha$ of both the WHAK and MAC models decreases at lower luminosities due to the $p$\ce{CO2}-dependence of weathering fluxes (see Figure \ref{fig:feedback_lum}). This causes higher susceptibility to both snowball and catastrophic warming events at lower luminosity, as the overall feedback resilience is lower; perturbations to the slow carbon cycle are less easily counteracted. Indeed, trials run with a lower fixed luminosity show a higher climate catastrophe rate. Trials run with a fixed luminosity of $L=0.9L_{\odot}$ show warming and snowball rates of 23\% and 6\% (Figure \ref{fig:lowL}), compared to 4.5\% and 0.2\% at $L=L_{\odot}$ (Section \ref{sec:tmax}).

\subsection{Model Limitations and Future Work}

Many aspects of the carbonate-silicate weathering feedback on Earth are ignored in our model for simplicity. Future improvements using our model might include some of the following.




\subsubsection{Climate Model}

If climate cools significantly, high latitude glaciation could locally shield silicate rocks from weathering, dependent on the local seasonality. This is a minor perturbation to global silicate weathering fluxes since most weathering occurs at low latitudes (e.g., \citealt{hartmann14}), but may act as a stabilizing feedback against the ice-albedo runaway.

Rainfall rates decrease not only with distance from the equator (Section \ref{sec:precip}), but also with distance inland from the coast \citep{Hayward1996}. River runoff is also required to travel long distances to the ocean without being evaporated first for carbonate precipitation. Distance from the ocean could thus also be a variable included in Equation \ref{eq:precip}.

$p$\ce{CO2} affects zonal temperature distribution (Equation \ref{eq:tlat}), with higher values of $p$\ce{CO2} flattening the equator-pole temperature gradient (e.g., \citealt{damste2010}).  This would likely reduce the effect of latitudinal drift on weathering fluxes in hot climates or lower solar luminosities, but exacerbate the effect in colder climates, leading to more snowball events. 

\subsubsection{Sulfide Oxidation and Silicate Weathering}

Recent analysis of solute-chemistry datasets from mountain streams suggest that \ce{CO2} drawdown due to chemical weathering is maximized at intermediate erosion rates \citep{bufe24}.  This is thought to be due to high erosion rates promoting sulfide oxidation, in which sulfide minerals (e.g., pyrite/\ce{FeS2}) are oxidized by oxygen gas or other oxidized species to \ce{H2SO4}/\ce{SO4^{2-}}. The resulting sulfuric acid/sulfate ions can then interact with carbonates to release \ce{CO2} gas into the atmosphere, offsetting \ce{CO2} drawdown from silicate weathering.  

However, significant \ce{CO2} release via sulfide oxidation likely requires a highly-oxidized atmosphere rich in \ce{O2} \citep{torres17}. Morever, high rates of sulfide oxidation may be mediated by biological activity \citep{boyd14}. In this study, we consider an abiotic atmosphere consisting of \ce{CO2}, \ce{H2O}, and an inert background gas (i.e., \ce{N2}).  Sulfide weathering would likely not be an important control of atmospheric \ce{CO2} on such worlds, but a more detailed future model could attempt to model its effects for oxygen-rich worlds. 

\subsubsection{Reverse Weathering and Groundwater Carbon}

Our model does not take into account `reverse weathering', where clays form on the seafloor and sequester cations from silicate weathering reactions, preventing the formation of carbonates and removal of \ce{CO2} from the atmosphere/ocean system (e.g., \citealt{dunlea17,isson18,trower19,krissansen20}).  This effect is not very pronounced in the slow carbon cycle of modern-day Earth due to creatures near the seafloor efficiently sequestering silica (\ce{SiO2}) to build their bodies. However, in an abiotic ocean, as assumed in this study, silica concentrations would be much higher, leading to more efficient clay formation. This effect would most directly affect our assumed $W_{\ce{CO2}}/W_{\ce{SiO2}}$ value, leading to lower \ce{CO2} sequestration fluxes.  Depending on the $p$\ce{CO2}/temperature dependencies of reverse weathering, which are poorly-constrained (see \citealt{krissansen20}), this may lead to either an additional stabilizing or destabilizing feedback in the slow carbon system.

Groundwater carbon fluxes could be another contributor to silicate weathering and could help explain the apparent mismatch (e.g., \citealt{Lee2019}) between estimates of volcanic \ce{CO2} and drawdown fluxes from silicate weathering (e.g., \citealt{zhang2020revisiting}).

\subsubsection{Glacial Impacts on Weathering}
Our model also neglects the potential impact of continental ice sheets on local silicate weathering fluxes. Under the extremely cold, dry conditions that characterize a hard Snowball state, global ice sheet coverage is hypothesized to shutdown (or at least severely throttle) continental silicate weathering rates, providing a mechanism for subsequent \ce{CO2} buildup and escape from glaciation \citep[e.g., ][]{Hoffman2017}. Some observations have suggested that the same is true locally on modern Earth; for example, Antarctica appears to contribute very little eroded/weathered material to the local waters \citep{edmond1973silica,hurd1977effect}. By this line of argument, it might appear that we overestimate global silicate weathering rates for any climate that would be cold enough for land glaciers to develop, neglecting a negative feedback on the carbon cycle whereby ice sheet growth would slow silicate weathering and stabilize the climate against further cooling. However, more recent observations and paleoclimate proxy records suggest that the net effect of land glaciers on global silicate weathering is less clear-cut. For example, ice sheet growth leads to falling sea level, exposing unconsolidated and highly weatherable shelf sediments to subaerial weathering, potentially enhancing CO$_2$ drawdown rates during glacial lowstands and increasing the amplitude of glacial-interglacial climate excursions \citep{wan2017enhanced}. Similarly, although cold temperatures reduce the intrinsic kinetic weathering rates from rocks underlying ice sheets, the simultaneous immense enhancement in erosion in glacial catchments can elevate rock surface area enough to compensate for or even dominate over the temperature effect, leading to similar weathering rates between glacial and non-glacial conditions \citep{munhoven1996glacial,anderson1997chemical,anderson2005glaciers,hatton2019investigation, li2022globally}. The precise interplay between silicate weathering and glaciation remains ambiguous and is likely not able to be represented adequately in a simple model like the one we present here.

\subsection{Mineralogical and Astrophysical Diversity}

Our model uses a single idealized chemical reaction (Reaction \ref{rxn:plag}) 
to represent a complex range of mineral compositions present on Earth.  Several values used in this study, including the solute concentration $C_{eq}$ and reaction rate $R_{n}$ (and their associated temperature and $p$\ce{CO2} dependencies) rely on this assumption. While previous studies have suggested that this reaction is a good first-order approximation of global weathering for Earth \citep{MAHER2011,Maher2014,Winnick2018}, including mineralogical diversity expected on Earth-like exoplanets \citep[e.g.,][]{hakim2021,Brantley2023} would paint a better picture of general habitability of rocky planets with global surface oceans.  However, this would likely require a large amount of laboratory work investigating the dissolution mechanics of more diverse mineralogy.

Our model currently only considers Earth-like planets orbiting a Sun-like star.  While we include varying stellar luminosity, the time evolution follows that of a Sun-like G star.  Given that M stars comprise the majority of stars in the galaxy and planets around quiet M stars are the most amenable to transmission and emission spectroscopy observations, especially with the James Webb Space Telescope (e.g., \citealt{kaltenegger09,de13,coy2025}), a logical next step could be implementing our model for Earth-like planets around M stars.  However, the atmospheric compositions of such terrestrial planets are not observationally constrained, and high extreme ultraviolet (XUV) radiation fluxes early in their host stars' lives may lead to complete loss of a secondary (outgassed) atmosphere \citep[e.g.,][]{zendejas2010,van2024,pass2025}. The tidally-locked nature of most (if not all) of these M star habitable zone planets would also introduce complex perturbations to the hydrologic cycle difficult to capture without 3D GCMs. Recent studies have proposed statistical frameworks to constrain where the inner edge of the habitable zone begins (e.g., \citealt{checlair19,schlecker24}), possibly giving observational evidence of the strength of the carbonate silicate weathering feedback on extrasolar worlds.  Future missions designed to investigate the atmospheres of Earth-twins such as the Habitable Worlds Observatory (HWO) will be required to carry out these feats.

\section{Conclusions and Implications for Habitability} \label{ch:conc}
We present a new idealized model for investigating the resilience of the carbonate-silicate weathering feedback of Earth-like planets around Sun-like stars based on the solute transport model presented in \citet{Maher2014} and \citet{Winnick2018}. 
Our idealized model's estimate for Earth's globally-averaged `climate feedback strength', which we define as the percentage change in weathering-induced \ce{CO2}-drawdown per increase in temperature, agrees with studies that consider the wide range of mineralogy and climate states present on Earth's surface.  We use this model to explore parameters relevant to the long-term climate stability of Earth-twins around Sun-like stars.

We find that the climate stability of planets with Earth-like plate tectonics and ocean depth is likely aided by more numerous continental plates, greater continental plate size, and more frequent continental plate collisions.  Worlds with few and small or slow continental plates are doomed to experience catastrophic warming from the build-up of \ce{CO2} as their surfaces become leached of minerals for weathering over time, decreasing their overall climate feedback strength.  Such worlds are also much more sensitive to small-scale perturbations in volcanic outgassing fluxes.  We find that additional sources of resupply for weatherable minerals--not just tectonic uplift spurred by collisions--are required for long-term climate stability, as Earth-like worlds with only tectonic uplift as a resupply mechanism for weatherable minerals show a $84\%$ climate catastrophe rate over 4 Gy.  
When approximating the effects of non-tectonic processes that aid weathering, such as dust transport and glacial erosion, the feedback is remarkably stabilizing, leading to a 95\% aggregate survival rate when omitting outgassing variability. This suggests that these non-tectonic processes may be essential \textit{in addition to} tectonic resurfacing for long-term climate stability on terrestrial exoplanets. When including simple stochastic volcanic outgassing variability (modeled after \citealt{baum2022snowball}), we find that long-term planetary habitability becomes highly unstable at low land fractions or long-term outgassing perturbations, suggesting that land fractions similar to, or larger than that of Modern-day Earth, are conducive for long-term habitability.  This conclusion is strengthened when considering the effects of solar luminosity increasing over time, 
i.e., the Faint Young Sun paradox.  We have also shown that seafloor weathering, based on $p$\ce{CO2} and temperature dependency estimates from previous studies \citep{sleep2001,Krissansen2018}, may have a much higher climate feedback strength than continental silicate weathering, and thus may be the dominant climate feedback on worlds with tectonic activity on the seafloor but little exposed land. 
Future field and laboratory work is needed to constrain the true $p$\ce{CO2} and temperature dependencies of seafloor weathering, as well as continental weathering at high $p$\ce{CO2} and temperatures relevant to rocky planets hotter than Earth.



\begin{acknowledgements}

B.P.C. and E.S.K. were supported by the Heising-Simons Foundation grant ``for the Simulation and Analysis of Initial Planetary States'' (grant \#2021-3119). R.J.G. received support from the National Aeronautics and Space Administration (NASA) through a contract with Oak Ridge Associated Universities (ORAU). The authors give special thanks to  Haitao Shang, Stilianos Louca, Rory Barnes, and two anonymous referees for their helpful comments in improving the manuscript. Figure \ref{fig:creservoir} was illustrated by Daniel Zhou at the University of Chicago. This work was completed in part with resources provided by the University of Chicago’s Research Computing Center.
\end{acknowledgements}

\software{
\texttt{cartopy} \citep{Cartopy}, \texttt{numpy} \citep{numpy}, \texttt{matplotlib} \citep{matplotlib},  \texttt{pandas} \citep{reback20}, \texttt{scipy} \citep{2020SciPy-NMeth}
}

\section*{Data Availability}
Python files for the \texttt{DISKWORLD} global weathering model and example scripts can be found on Zenodo \citep{coy2025_zenodo} \dataset[DOI: 10.5281/zenodo.12668348]{https://doi.org/10.5281/zenodo.12668348}. All additional data is available upon request.

\bibliographystyle{aasjournal7}
\bibliography{main}

\appendix

\section{Additional Model Methods}
\restartappendixnumbering 

\subsection{MAC Model Temperature and $p$\ce{CO2} Dependencies} \label{sec:tdepend}

In MAC \citep{Maher2014}, $k_{eff}$ has an Arrhenius dependence on temperature:

\begin{equation} \label{eq:MACtemp2}
    \frac{k_{eff}(T)}{k_{eff}(T_{ref})}=e^{\frac{E_{a}}{R}\left(\frac{1}{T_{ref}}-\frac{1}{T}\right)}
\end{equation}
where $E_{a}$ is the apparent activation energy of mineral dissolution and $R$ is the gas constant. Using values presented in MAC, this effect causes $k_{eff}$ to change by $\sim70\%$ for a \SI{10}{\celsius} change in temperature. Systems including these temperature dependencies are more resilient to changes in weathering flux since increases in $k_{eff}$ also increase $Dw$ (Equation \ref{eq:dw}), so we include the temperature dependency of $k_{eff}$ in our model.  Due to this effect, weathering fluxes using a soil age for mountains (\SI{8e3}{yr}) at an initial temperature of $T=\SI{15}{\celsius}$ increase by $\sim30\%$ for a \SI{10}{\celsius} increase.

\citet{Winnick2018} show that $C_{eq}$ depends on soil $p$\ce{CO2}; for an example dissolution reaction where a primary mineral M reacts with carbonic acid to form a secondary mineral S and cation weathering products A and B: 

\begin{equation}
    \ce{$m$M + 2CO2(g) + H2O <-> $s$S + 2HCO3- + $a$A + $b$B + $c$SiO2(aq)}
\end{equation}
the equilibrium constant is:
\begin{equation}
    K=\frac{\ce{[HCO3^{-}]^{2}}\ce{[A]^{a}}\ce{[B]^{b}}\ce{[SiO2]^{c}}}{\ce{(pCO2)}^{2}}.
\end{equation}
Assuming stoichiometry sets the relative concentration of dissolution products in equilibrium, the $p$\ce{CO2} dependence of dissolution products is,

\begin{equation}
    \ce{[SiO2]}\propto\ce{(pCO2)}^{\frac{2}{2+a+b+c}}
\end{equation}

We use the stoichiometry of the dissolution of plagioclase feldspar (An20) and the precipitation of halloysite to obtain:
\begin{equation}\label{eq:pco2dep2}
    C_{eq}(p\textrm{CO$_2$})=C_{eq}(p\textrm{CO$_{2,\textrm{\,ref}}$})\left(\frac{p\ce{CO2}}{p\textrm{CO$_{2,\textrm{\,ref}}$}}\right)^{0.316}.
\end{equation}
Here $p\textrm{CO$_{2,\textrm{\,ref}}$}=\SI{380}{\micro bar}$ \citep{MAHER2011}. This formulation assumes that weathering occurs primarily in an open-system, i.e. the $p$\ce{CO2} of waters is always in equilibrium with atmospheric/soil $p$\ce{CO2}. \citet{Winnick2018} show that the open-system approximation using the reaction for the dissolution of plagioclase feldspar (An20) to halloysite is in good agreement with field observations from \citet{IBARRA2016}.

\begin{figure}
    \centering
    \includegraphics[width=.8\linewidth]{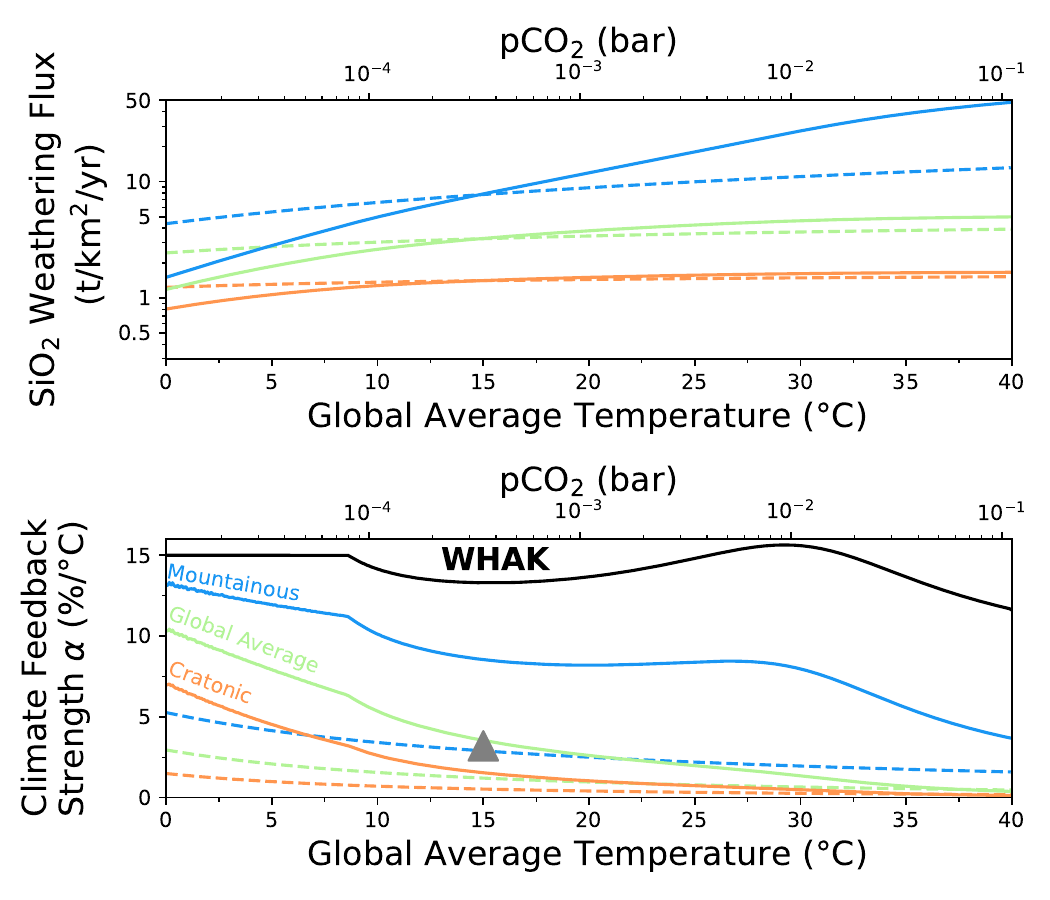}
    \caption{
    (Top) \ce{SiO2} weathering flux as a function of globally averaged surface temperature for various soil ages following Figure \ref{fig:feedback}. Solid lines include the temperature and $p$\ce{CO2} dependencies derived in \citet{Winnick2018} assuming soil $p$\ce{CO2} is in equilibrium with atmospheric $p$\ce{CO2} (Equations \ref{eq:MACtemp} and \ref{eq:pco2dep}), while dashed lines represent the original formulation of \citet{Maher2014}. 
    (Bottom) Climate feedback strength $\alpha$ as a function of global temperature following Figure \ref{fig:feedback}.}
    \label{fig:feedback_original}
\end{figure}

This effect strengthens the feedback--as $p$\ce{CO2} rises, a higher equilibrium solute concentration $C_{eq}$, while lowering $Dw$, will lead to larger weathering flux values (Equation \ref{eq:MACw}) and increase \ce{CO2} drawdown.  A comparison of our model's weathering fluxes and the original model presented in \citet{Maher2014} is shown in Figure \ref{fig:feedback_original}.

\begin{figure}
    \centering
    \includegraphics[width=.4\linewidth]{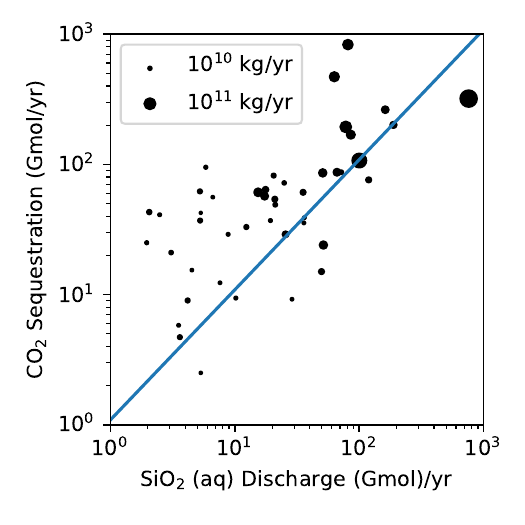}
    \caption{Silica discharge rates and carbon consumption rates associated with various rivers presented in \citet{GAILLARDET1999}. The size of each point represents the total discharge of dissolved solids and associated weights for each river in our slope determination. The line represents the linear relationship of slope 1.087 used in our model.}
    \label{fig:co2sio2}
\end{figure}

\subsection{Euler Poles} \label{sec:ap_euler} 
Plates in our simulation rotate around randomized Euler poles at fixed velocities.  
This is accomplished by first choosing a randomized longitude and latitude ($\Lambda, \Phi$) for each disk's rotational (Euler) pole. Latitudes are chosen so that they are equiareally spaced. All positions are calculated in the new Euler frame ($\lambda_{e},\phi_{e}$) and subsequently rotated into the planetary frame ($\lambda_{p},\phi_{p}$).  

Rotation matrices are defined as 
\begin{equation}
    \mathbf{R}=\mathbf{A}\mathbf{B}\mathbf{C}
\end{equation}
where 
\begin{equation}
    \mathbf{A}=\begin{vmatrix}
\cos{\alpha} & -\sin{\alpha} & 0 \\
\sin{\alpha} & \cos{\alpha} & 0\\
0 & 0 & 1
\end{vmatrix},
\end{equation}
$\alpha=\Lambda+\frac{\pi}{2}$,

\begin{equation}
    \mathbf{B}=\begin{vmatrix}
1 & 0 & 0 \\
0 & \cos{\beta} & -\sin{\beta}\\
0 & \sin{\beta} & \cos{\beta}
\end{vmatrix},
\end{equation}
$\beta=\frac{\pi}{2}-\Phi$, and

\begin{equation}
    \mathbf{C}=\begin{vmatrix}
\cos{(-\alpha)} & -\sin{(-\alpha)} & 0 \\
\sin{(-\alpha)} & \cos{(-\alpha)} & 0\\
0 & 0 & 1
\end{vmatrix}.
\end{equation}
Rotation between the Euler and planetary frame is performed via
\begin{equation}
    \vec{x}_{p}\mathbf{R}=\vec{X}_{e},
\end{equation}
where $\vec{x}$ is the ($x,y,z$) coordinates of the disk center, which can then be converted to ($\lambda,\phi$) coordinates.

\begin{figure}
    \centering
    \includegraphics[width=0.6\linewidth]{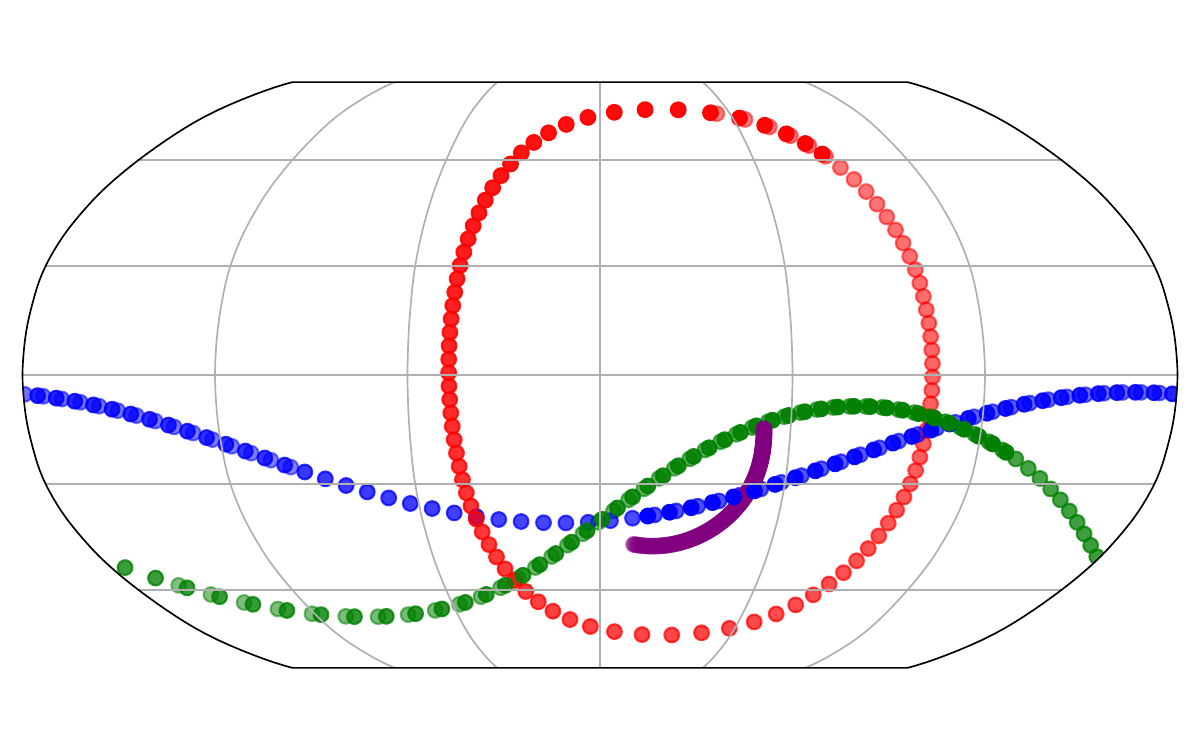}
    \caption{Example trajectories of four disk centers over 4 Gy, assuming a initial (and constant) velocity distribution of $V=4\pm\SI{2}{cm.yr^{-1}}$. }
    \label{fig:trajectory}
\end{figure}

To define a starting location, another latitude and longitude point is then chosen in the Euler frame.  Disk centers rotate around the Euler pole at a constant latitude (in the Euler frame, i.e. $\frac{d\lambda_{e}}{dt}=\Omega, \frac{d\phi_{e}}{dt}=0$).  The velocity ($\Omega=\frac{v}{2\pi R_{pl}\cos{\phi_{e}}}$) of this rotation is determined by a Gaussian distribution ($v=V\pm\sigma_{V}$, with parameters in Table \ref{tab:parameters}).  Example disk trajectories are shown in Figure \ref{fig:trajectory}.

\subsection{Carbon Partitioning and Ocean Chemistry} \label{sec:csys}

To calculate partitioning of inorganic carbon within the ocean and atmosphere, we use a modified version of the software package \texttt{csys} \citep{Zeebe2001}. In \texttt{csys}, dissolved inorganic carbon (DIC) is partitioned between \ce{H2CO3}/\ce{CO2(aq)}, \ce{HCO3^-}, and \ce{CO3^2-}, governed by the following reactions:

\begin{equation}
    \ce{CO2(g) <-> CO2(aq)}.
\end{equation}

\begin{equation}
    \ce{H2O + CO2(aq) <-> H2CO3 <-> H+ + HCO3-}.
\end{equation}

\begin{equation}
    \ce{HCO3- <-> H+ + CO3^{2-}}.
\end{equation}

The equilibrium constants for these reactions ($K_{H}$, $K_{1}$, and $K_{2}$, respectively) depend on ocean temperature, salinity, and atmospheric pressure \citep{Zeebe2001}. We assume an ocean salinity of \SI{35}{g.L^{-1}}.

\begin{figure}
    \centering
    \includegraphics[width=0.95\linewidth]{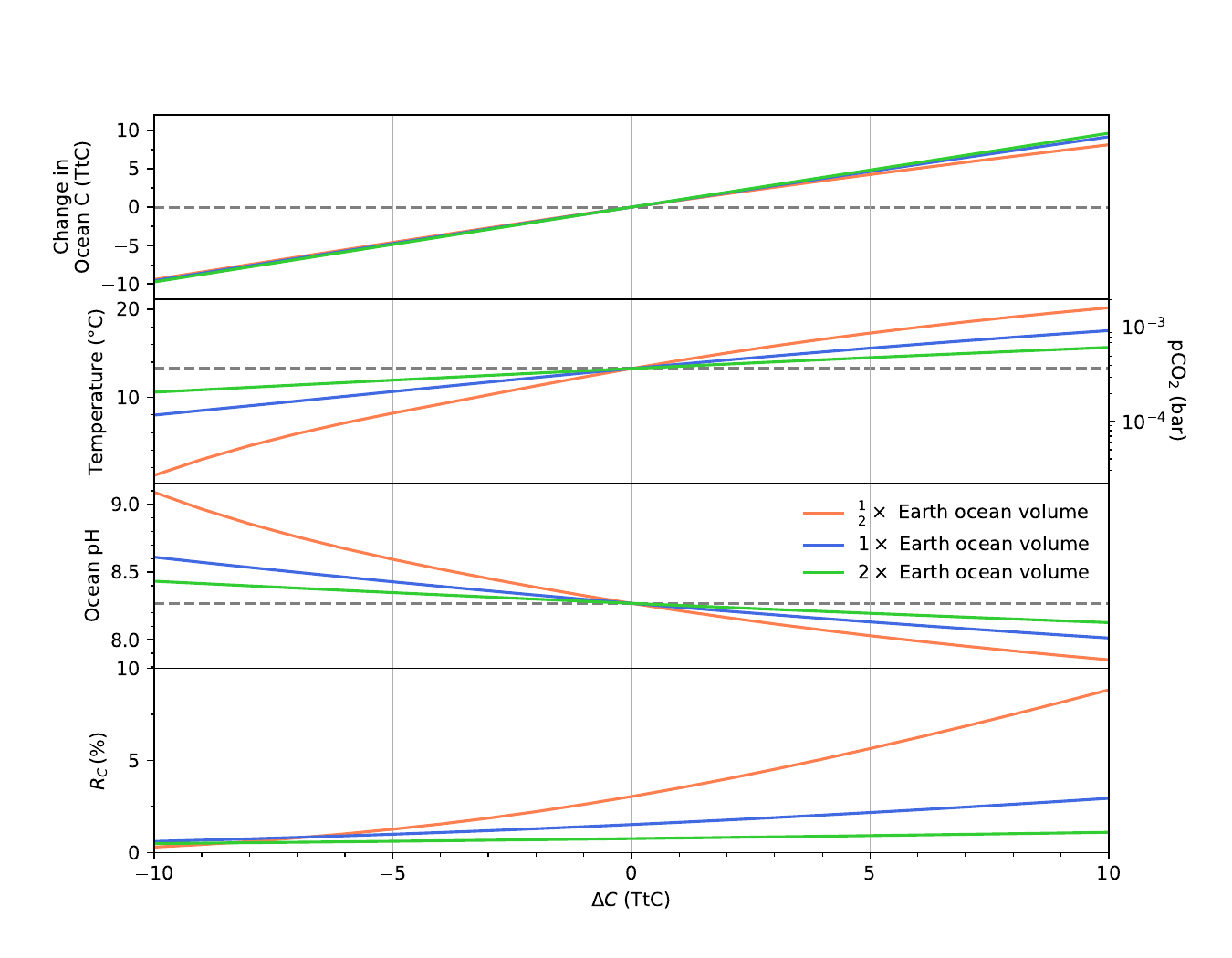}
    \caption{Effects of ocean volume on carbon ocean-atmosphere carbon partitioning in our model; long-term (timescale$\gg5000$ yr) response of atmospheric/ocean \ce{CO2} content, globally average surface temperature, and ocean pH to carbon perturbations up to 10 TtC [teratonnes of carbon ($\SI{1}{TtC}=\SI{1000}{GtC}$)], 10 TtC is about $25\%$ of the current ocean-atmosphere reservoir, based on methods outlined in Section \ref{sec:csys}.  Blue lines represent an Earth-like ocean volume, whereas green and orange represent half and twice that volume, respectively.  $R_{C}$ is the ratio of carbon stored in the atmosphere to to carbon stored in the ocean. Dashed lines represent initial conditions.  Initial states were created assuming a pCO$_2$ of \SI{280}{\micro bar}, resulting in $\sim\frac{1}{2}$ and $\sim2$ times the total carbon reservoir for the shallow and deep ocean scenarios, respectively. The climates of planets with smaller oceans are more sensitive to a given \ce{CO2} perturbation.   Increasing $R_{C}$ values for \ce{CO2} injections show that the ocean carbon sink becomes less effective as more carbon is added to the system.}
    \label{fig:dC}
\end{figure}

As carbon is added to the ocean-atmosphere system, the ocean becomes more acidic and the fraction of carbon stored in the atmosphere versus ocean increases (see Figure \ref{fig:dC}). Ocean surface temperature also affects the solubility of \ce{CO2}, with higher temperatures increasing the fraction of carbon stored in the atmosphere. For these reasons, in addition to $p$\ce{CO2}, the amount of DIC stored in the ocean, $C_{ocean}$, and the pH of the ocean are kept track of in our simulations.

Carbon is partitioned between the atmosphere and ocean following \cite{graham21}. They assume that calcium carbonate saturation dictates atmosphere-ocean partitioning on multi-millennial timescales, i.e. the calcium carbonate saturation state,
\begin{equation}\label{eq:sat}
    \Omega=\frac{\ce{[Ca^{2+}]}\ce{[CO3^{2-}]}}{K_{sp}}=1,
\end{equation}
where $K_{sp}$ is the solubility product of calcite and is a function of temperature, salinity, and pressure.  A net zero charge balance constraint of 
\begin{equation}
    \ce{2[Ca^{2+}] + [H^{+}]=[HCO3^{-}] + 2[CO3^{2-}] + [OH-]}
\end{equation}
allows for solving for \ce{[H^{+}]} as a function of $p$\ce{CO2} and temperature via the fourth-degree polynomial,
\begin{equation}
    2K_{sp}\ce{[H+]}^{4}+\frac{K_{1}K_{2}}{K_{H}}p\ce{CO2}\ce{[H+]}^{3}-\left(\frac{K_{1}K_{2}}{K_{H}^2}p\ce{CO2}^{2}+\frac{K_{1}K_{2}}{K_{H}}K_{w}p\ce{CO2}\right)\ce{[H+]}-2\frac{K_{1}^{2}K_{2}^{2}}{K_{H}^{2}}p\ce{CO2}^{2}=0,
\end{equation}
where $K_{w}$ is the (temperature- and pressure-dependent) ion-product constant of water. The resulting pH and $p$\ce{CO2} from a given addition of carbon is solved for via a minimization scheme that compares the total atmospheric and ocean carbon to the amount of added carbon.  To enhance computational efficiency, the relevant equilibrium constants ($K_{1}$, $K_{2}$, $K_{H}$, $K_{w}$, and $K_{sp}$) are calculated in a 500$\times$100 grid in $\log_{10}(p\ce{CO2})$ and luminosity space (spanning -8 to 2 and 0.7 to 1, respectively) and linearly interpolated. This results in $<0.1\%$ error for equilibrium constants over the vast majority of $p$\ce{CO2}/luminosity values within this range.  This approach results in a slightly larger ocean carbon reservoir (41 TtC) than typically estimated for modern-day earth (38 TtC, \citealt{Zeebe2012,Lee2019}), however recreates modern-day ocean pH estimates (8.1) when $p\ce{CO2}=\SI{420}{\micro bar}$.

We do not latitudinally resolve ocean temperatures in our model, and set the ocean temperature equal to the globally-averaged surface temperature $\overline{T}$. \ce{Ca^{2+}} is a \textit{conservative} ion, meaning that its concentration is generally unperturbed by changes in ocean temperature, salinity, and pressure. 
In our model, we assume that precipitation is efficient at $\Omega=1$, but in an abiotic ocean where calcium carbonate precipitation is non-biogenic \citep[sometimes termed a ``Strangelove ocean''; ][]{Zeebe2003}, it is possible that the ocean would need to reach a much higher saturation state than $\Omega=1$ to trigger widespread carbonate precipitation. Experimental studies have suggested critical $\Omega$ values for abiotic precipitation ranging from $\gg20$ to $>100$ \citep[e.g., ][]{lee2010influences,strauss2020mineralogical}. A higher critical $\Omega$ would force the ocean to achieve larger concentrations of $\ce{[Ca^{2+}]}$ and $\ce{[CO3^{2-}]}$ in order to drive carbonate precipitation. The resulting system would require a higher oceanic pH and higher carbonate ion concentration to sustain a given rate of CO$_2$ drawdown, implying a greater capacity to buffer the climate against transient changes in weathering or outgassing fluxes, since flux changes of a given size would translate into proportionally smaller changes in the stock of molecules in the atmosphere/ocean system \citep{honing2020impact}.

For trials where we vary the total amount of land, we assume a constant ocean depth, i.e. the total ocean volume scales inversely with the amount of land area.  To solve for \ce{CO2} partitioning for various ocean volumes, we assume that [\ce{Ca2+}] (and thus starting \ce{pH} and DIC concentration) is constant.  For a given $p$\ce{CO2}, this leads to a higher total \ce{CO2} ocean reservoir that scales directly with ocean volume.  Larger oceans (see Section \ref{sec:ap_ocean_depth}) are generally better buffers of \ce{CO2} injections, owing to the fact that a given amount of \ce{CO2} will have less effect on the DIC \textit{concentration}.  Differences in ocean \ce{CO2} buffering can be seen in Figure \ref{fig:dC}.

\subsection{Closeness Factor}\label{sec:closeness}

To smooth out numerical artifacts associated with using a finite number of grid points, we introduce a `closeness factor' $f$. $f$ is based on the distance of the colliding point to the center of the nearest disk, $d_{min}$:

\begin{equation}
f=
    \begin{cases}
        1 & \text{if $d_{min}\leq0.8 R$}\\
        \frac{1-d_{min}}{0.2} & \text{if $d_{min}>0.8$}.
    \end{cases}
\end{equation}
$f$ is set to 0 in non-colliding points. $f$ affects how local weathering fluxes are calculated:
\begin{equation}\label{eq:collidingf}
W_{i}=
    \begin{cases}
        fW(\tau_{s}=\tau_{s,mountain})+(1-f)W(\tau_{s}=\tau_{s,max}/2) & \text{if colliding}\\
        W(\tau_{s}=\tau_{s,i}) & \text{if not colliding}.
    \end{cases}
\end{equation}
This causes glancing collisions to retain older effective soil ages, while more deeply penetrating collisions completely reset the soil age.

\section{Sensitivity Tests}\label{ap:sensitivity}
In this section, we explore the effects of varying tectonic parameters not discussed in the main text.  All trials vary $A_{land}$ and $N_{disk}$ as discussed in Section \ref{sec:params} and are based on the model used in Section \ref{sec:tmax}.
\subsection{Disk Velocity}

To test the effects of changing disk velocity $V$, which we prescribe as $4\pm\SI{2}{cm.yr^{-1}}$ in our main trials, we run two additional scenarios using velocity spreads of $V=40\pm\SI{20}{cm.yr^{-1}}$ (`fast' case) and $V=0.4\pm\SI{0.2}{cm.yr^{-1}}$ (`slow' case).

In the slow case, initial disk configurations (whether or not a disk is colliding) greatly affect survival.  Initial configurations lacking any colliding disks have a much lower chance to experience a collision and are essentially doomed to die on Gyr timescales. Even in systems that are colliding, the distribution of colliding and not colliding area does not change much; the non-colliding regions will age while the colliding regions will stay at the prescribed soil age (\SI{8}{kyr}), resulting in a slow death.  Even though there are a high amount of survivors ($\sim99\%$), they are (almost) all on a slow path to death, driven by rising $p$\ce{CO2}. Increasing disk velocity greatly increases the variability of collisions, disk latitude, and weatherability.  This causes most trials to have an extremely high variability in atmospheric $p$\ce{CO2}, and several end via catastrophic warming.

\begin{figure}[htp]
\centering
\subfloat[$V=0.4\pm\SI{0.2}{cm.yr^{-1}}$]{
  \includegraphics[clip,width=0.75\columnwidth]{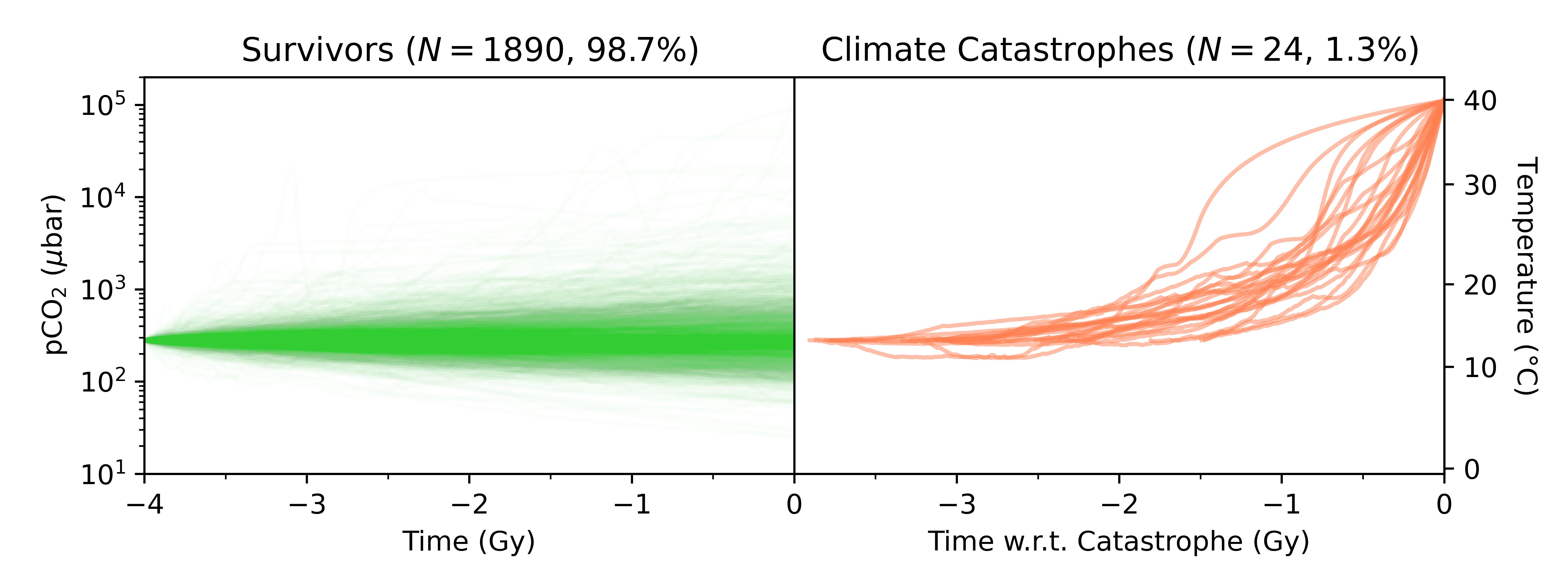}
  }

\subfloat[$V=40\pm\SI{20}{cm.yr^{-1}}$]{
  \includegraphics[clip,width=0.75\columnwidth]{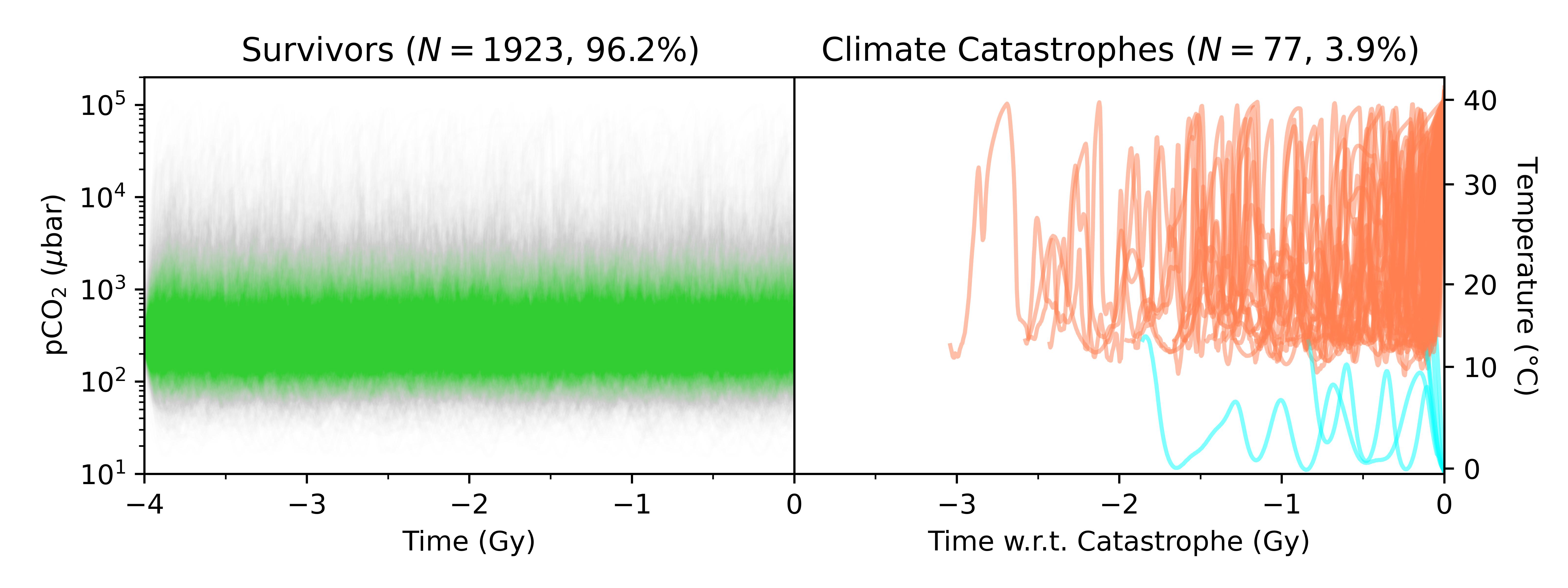}
  }

\caption{$p$\ce{CO2} curves for 4000 trials using a $10\times$ (a) slower and (b) faster average disk velocity than our base trials. While both trials show relatively high survival rates, initial collision states (disk configurations) determine long-term behavior for slower disk velocities, whereas faster velocities introduce more variability in $p$\ce{CO2} and higher collision frequency.}

\end{figure}

\subsection{Silicate Weathering Activation Energy}

Our model assumes an apparent activation energy of mineral dissolution of \SI{38}{kJ/mol}. The exact value that should be used is still a matter of debate; for example \citet{Krissansen2017} find a best fit of $E_{a}=\SI{20}{kJ.mol^{-1}}$ through their Markov-Chain Monte Carlo (MCMC) model, whereas Graham and Pierrehumbert use an effective value of \SI{62}{kJ.mol^{-1}}.  \citet{Krissansen2017} further argue that, while most studies assume a value of $E_{a}=50\sim\SI{100}{kJ.mol^{-1}}$, that there is a 98\% chance based on proxy evidence that this value should be $<\SI{45}{kJ.mol^{-1}}$.  It is important to note that most of these models calibrate this value based on the WHAK and not the MAC model and thus might not be translatable to our model framework.

\begin{figure}
    \centering
    \includegraphics[width=0.6\linewidth]{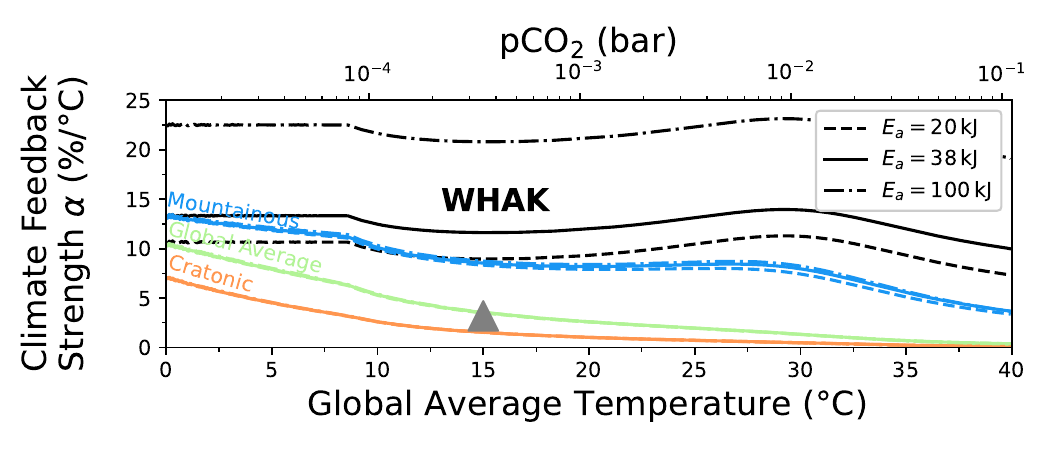}
    \caption{Climate feedback strength $\alpha$ as a function of globally averaged temperature for varying effective soil age and activation energy of mineral dissolution $E_{a}$, similar to Figure \ref{fig:feedback}.  The gray triangle represents the global estimate from \citet{Brantley2023}.}
    \label{fig:sensitivity_E_act}
\end{figure}

The MAC model is much more reliant on other parameters, such as the effective soil age, river runoff rate, and partial pressure of $p$\ce{CO2}, than the WHAK model \citep{Graham2020}. The value assumed for $E_{a}$ (used in Eq. \ref{eq:MACtemp2}) has a relatively small effect on climate feedback strength, as shown in Figure \ref{fig:sensitivity_E_act}. 
We ran additional sensitivity tests adjusting $E_{a}$ over $20-\SI{100}{kJ.mol^{-1}}$ and found no significant difference in overall climate behavior or survival rates.

\subsection{Ocean Depth} \label{sec:ap_ocean_depth}

\begin{figure}[htp]
\centering
\subfloat[$\frac{1}{2}\times$ ocean depth]{
  \includegraphics[clip,width=0.8\columnwidth]{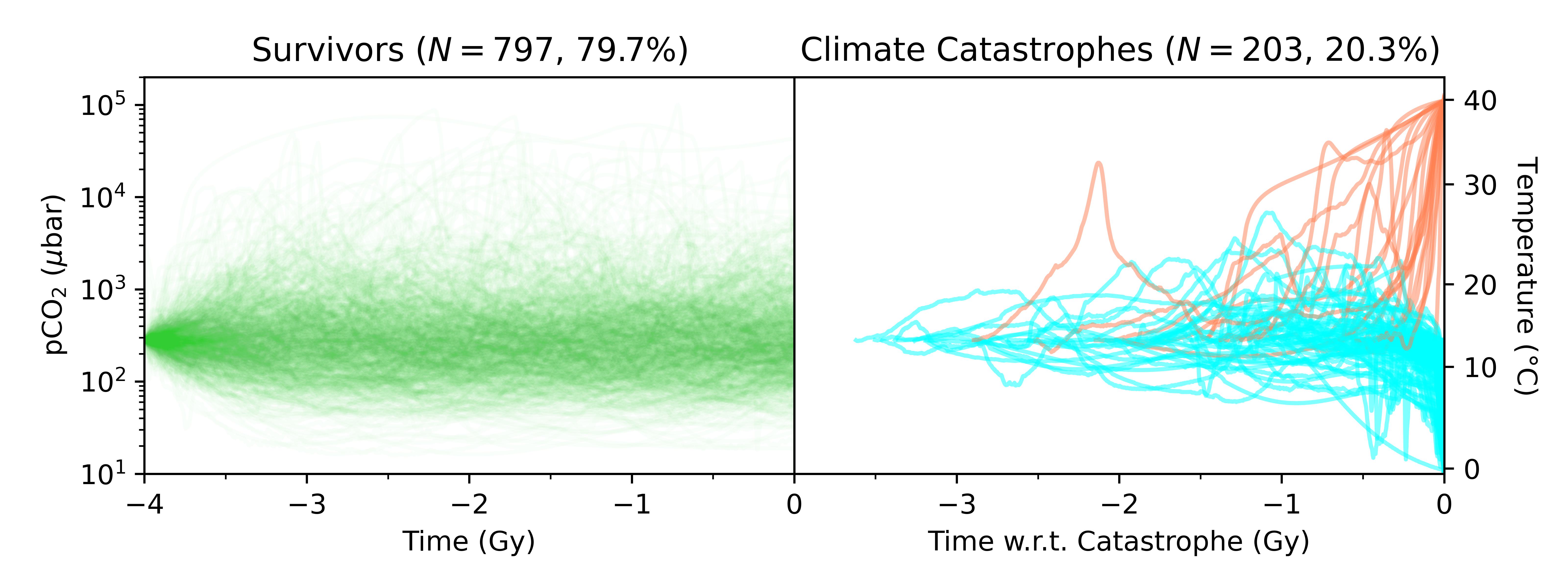}
  }

\subfloat[$2\times$ ocean depth]{
  \includegraphics[clip,width=0.8\columnwidth]{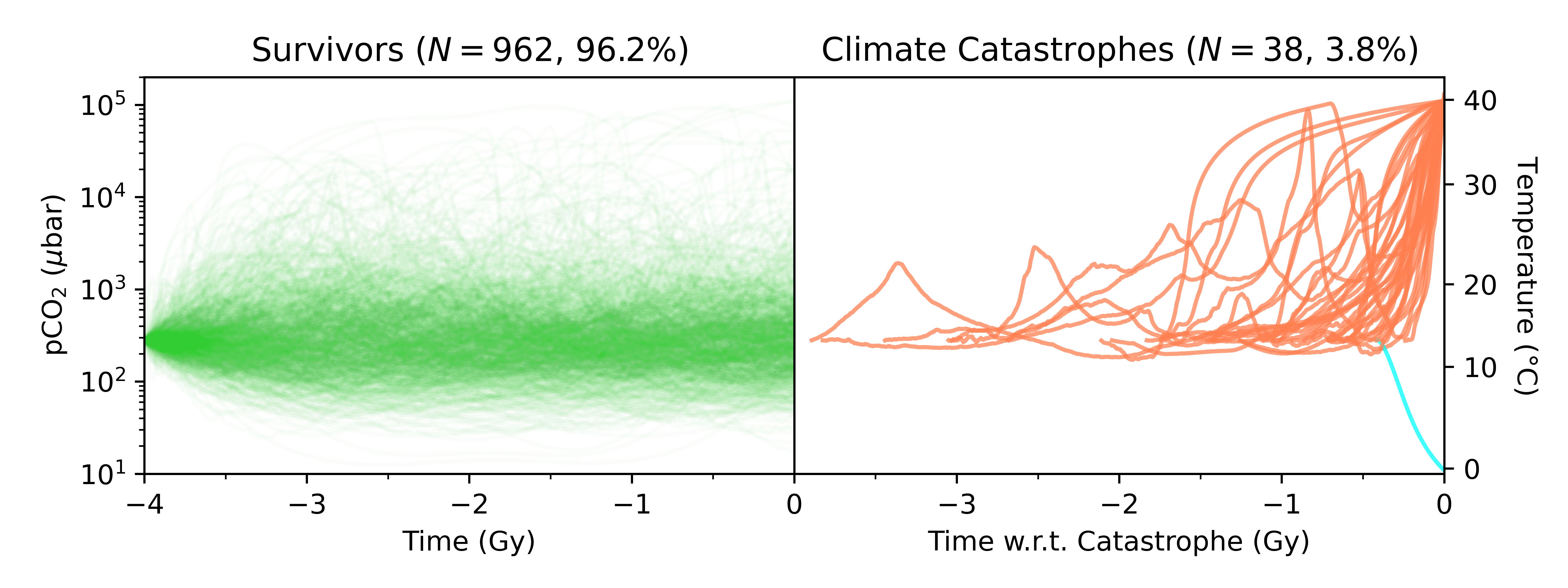}
  }

\caption{Model output $p$\ce{CO2} for 2000 trials with (a) half of Earth's current ocean depth and (b) twice current ocean depth. As suggested in Figure \ref{fig:dC}, larger ocean volumes act as a stronger buffer to perturbations in the slow carbon cycle and show lower climate catastrophe rates (3.8\% vs. 20.3\%}). \label{fig:varocean}

\end{figure}

As shown in Appendix \ref{sec:csys}, the volume of a planet's ocean greatly affects its susceptibility to perturbations in \ce{CO2}, with larger oceans reducing susceptibility by acting as a buffer for the atmosphere.  We test the effects of changing our average ocean depths by $\frac{1}{2}\times$ and $2\times$, corresponding to roughly \SI{2}{km} and \SI{8}{km}, respectively. These depths are inherently tied to the land fraction, but are likely not deep enough to drown all land.  Trial results can be seen in \ref{fig:varocean}.  As expected, a deeper ocean acts a stronger buffer to perturbations in the slow carbon cycle, leading to more stable climate on geologic timescales, whereas shallow oceans are particularly susceptible to snowball events.

\subsection{Artificially Increasing Weathering Fluxes} \label{sec:ap_fudge}

As noted in \citet{baum2022b} and mentioned in the Supplementary Material of \citet{Maher2014}, the MAC model tends to underestimate global \ce{CO2} drawdown due to continental weathering by a factor of $\sim 3$ when compared to estimates for modern-day weathering-induced \ce{CO2} drawdown.   To compensate for this, \citet{baum2022b} use a much higher $C_{eq}$ value and increase the default flow path length from \SI{1}{m} to \SI{1.5}{m}, which roughly reproduces the correct global fluxes.  Indeed, we find that following their methodology leads to higher global weathering fluxes comparable to modern-day estimates for Earth.  However, they accomplish this by using the equilibrium concentration of [\ce{HCO3-}] (for the same Reaction \ref{rxn:plag}) given in \citet{Winnick2018}, where all constants in \citet{Maher2014} are calibrated for \ce{SiO2} weathering fluxes. This results in an unrealistic $22\times$ larger value for the thermodynamic maximum concentration of \ce{SiO2} ($C_{eq})$.

\begin{figure}
    \centering
    \includegraphics[width=0.9\linewidth]{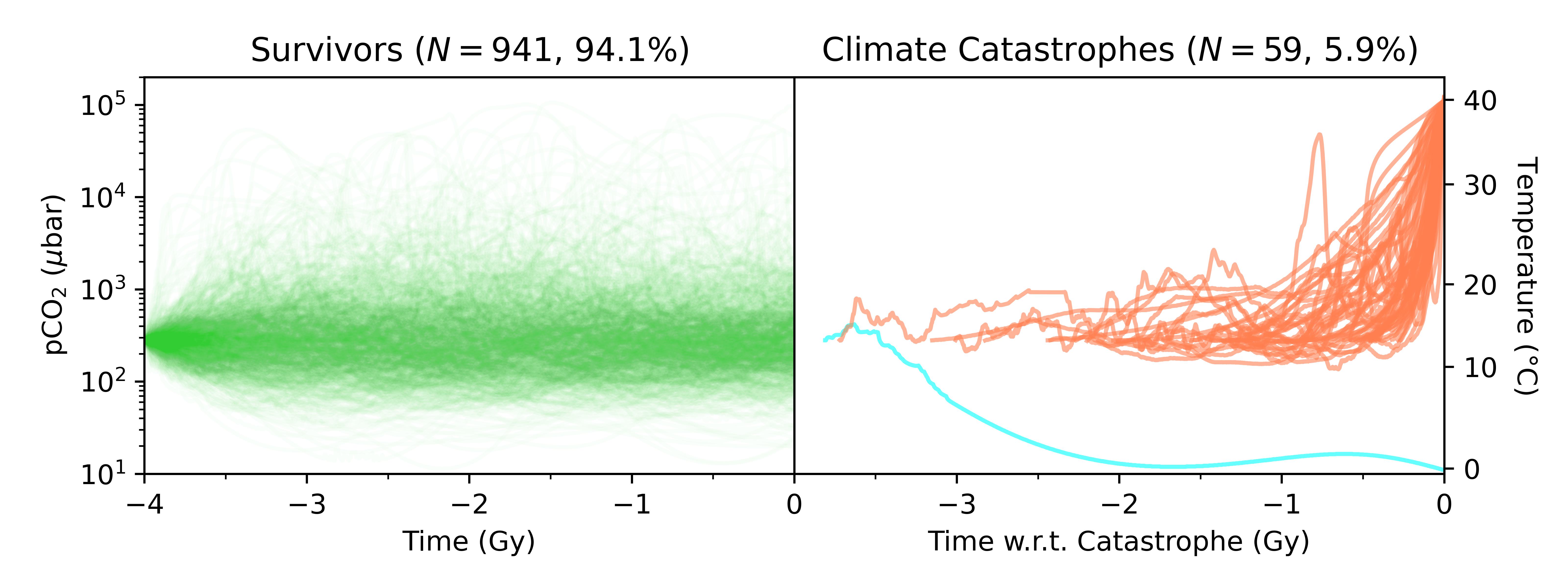}
    \caption{Model output for 1000 trials including artificially increased weathering fluxes, as described in Appendix \ref{sec:ap_fudge}.  Runs show similar behavior and catastrophe rates to results shown in Section \ref{sec:tmax} (Figure \ref{fig:wmaxt}).}
    \label{fig:fudge}
\end{figure}

We investigate the effects of marginally increasing the \textit{continental} weathering flux predicted by Equation \ref{eq:MACw} by $3\times$ so that our weathering fluxes roughly match that of modern-day Earth.  
We find that these trials show roughly the same behavior as our base trials, with similar climate catastrophe rates (Figure \ref{fig:fudge})--catastrophic warming rates of 5.8\% vs. 4.5\% (base) and snowball rates of 0.1\% vs. 0.2\%, reaffirming our conclusions.

\end{document}